\documentclass[structabstract]{aa}  
\usepackage{graphicx}
\usepackage{txfonts}
\usepackage{natbib}
\bibpunct{(}{)}{;}{a}{}{,}
\def\spose#1{\hbox to 0pt{#1\hss}}
\def\ltsimm{\mathrel{\spose{\lower 3pt\hbox{$\sim$}}
        \raise 2.0pt\hbox{$<$}}}
\def\gtsimm{\mathrel{\spose{\lower 3pt\hbox{$\sim$}}
        \raise 2.0pt\hbox{$>$}}}
\def\aap{{\rm A\&A}}
\def\apj{{\rm ApJ}}
\def\apjs{{\rm ApJS}}
\def\aj{{\rm AJ}}
\def\mnras{{\rm MNRAS}}

\def\apss{{\rm Ap\&SS}}
\def\araa{{\rm ARA\&A}}

\def\cpc{{\rm Comp.~Phys.~Comm.}}

\begin{document}
   \title{Investigating the X-ray emission from the massive WR+O binary WR~22 using 3D hydrodynamical models}
   \titlerunning{WR~22: Hydrodynamical modelling and X-ray emission}


   \author{E.~R.~Parkin\inst{1,2} \and E.~Gosset\inst{1}\thanks{F.~R.~S.-FNRS (Belgium)} }

   \institute{Institut d'Astrophysique et de
         G\'{e}ophysique, Universit\'{e} de Li\`{e}ge, 17, All\'{e}e
         du 6 Ao\^{u}t, B5c, B-4000 Sart Tilman, Belgium \\
        \and 
         Research School of
       Astronomy and Astrophysics, Mount Stromlo Observatory,
       Australian National University, Cotter Road, Weston Creek, ACT
       2611, Australia \\
       \email{parkin@mso.anu.edu.au}
   }

   \date{Received 24/02/2011; accepted 25/03/2011}

 
  \abstract
      {}
      {We examine the dependence of the
      wind-wind collision and subsequent X-ray emission from the
      massive WR+O star binary WR~22 on the acceleration of the stellar
      winds, radiative cooling, and orbital motion.}
      {Three dimensional (3D) adaptive-mesh
      refinement (AMR) simulations are presented that include
      radiative driving, gravity, optically-thin radiative cooling,
      and orbital motion. Simulations were performed with
      instantaneously accelerated and radiatively driven stellar
      winds. Radiative transfer calculations were performed on the
      simulation output to generate synthetic X-ray data, which are
      used to conduct a detailed comparison against observations.}
      {When instantaneously accelerated
      stellar winds are adopted in the simulation, a stable wind-wind
      collision region (WCR) is established at all orbital phases. In
      contrast, when the stellar winds are radiatively driven, and
      thus the acceleration regions of the winds are accounted for,
      the WCR is far more unstable. As the stars approach periastron,
      the ram pressure of the WR's wind overwhelms the O star's and,
      following a significant disruption of the shocks by non-linear
      thin-shell instabilities (NTSIs), the WCR collapses onto the O
      star. X-ray calculations reveal that when a stable WCR exists
      the models over-predict the observed X-ray flux by more than two
      orders of magnitude. The collapse of the WCR onto the O star
      substantially reduces the discrepancy in the $2-10\;$keV flux to
      a factor of $\simeq 6$ at $\phi=0.994$. However, the observed
      spectrum is not well matched by the models.}
      {We conclude that the agreement between the models and
        observations could be improved by increasing the ratio of the
        mass-loss rates in favour of the WR star to the extent that a
        normal wind ram pressure balance does not occur at any orbital
        phase, potentially leading to a sustained collapse of the WCR
        onto the O star. Radiative braking may then play a
        significant r\^{o}le for the WCR dynamics and resulting X-ray
        emission.}


   \maketitle
%

\section{Introduction}
\label{sec:intro}
   Hot luminous massive stars possess powerful radiatively driven
   stellar winds \citep[for a recent review see][]{Puls:2008}. In a
   binary system consisting of two such stars, the collision of the
   winds generates a region of high temperature ($T > 10^7\;$K) plasma
   which emits prolifically at X-ray wavelengths
   \citep[][]{Prilutskii:1976, Cherepashchuk:1976, Luo:1990,
     Stevens:1992}. Depending on the parameters of the winds and the
   orbit, the dynamics of the postshock gas in the WCR can cover a
   diverse range \citep{Stevens:1992}. For instance, in long-period
   binaries (i.e. of the order of years) the postshock gas is expected
   to be quasi-adiabatic for the most part, whereas in short-period
   (i.e. a few days) systems the postshock gas is expected to be
   highly radiative. As a result, eccentric intermediate period systems (of
   the order of 100's of days), such as WR22 ($e \simeq0.559$,
   $P\simeq 80\;$days - see Tables~\ref{tab:system_parameters} and
   \ref{tab:stellar_parameters}), provide the prospect of
   transitioning between these two extremes. Such a transition in the
   state of the postshock gas has also been found in simulations of
   shorter period eccentric OB star systems by \cite{Pittard:1998b}
   and \cite{Pittard:2009}.

   The X-ray emission from a colliding winds binary system acts as a
   direct observational probe of the postshock winds, and hence an
   indirect probe of the preshock winds \citep[e.g.][]{Stevens:1996,
     Pittard:2002, DeBecker:2006}. Recently, an analysis of {\it
     XMM-Newton} observations of WR22 by
   \cite*{Gosset:2009}~\citep[hereafter][]{Gosset:2009} characterised
   the X-ray emission using a two-component spectrum consisting of a
   soft component at $~\sim0.6\;$keV and a harder component at $\sim
   2-4.5\;$keV. However, difficulties were encountered as wind-wind
   collision models were found to over-predict the observed X-ray flux
   by more than two orders of magnitude.

   Considering the separation of the stars in WR22
   (Table~\ref{tab:system_parameters}), and the dominant WR wind ram
   pressure characteristic of WR+O binary systems
   \citep[e.g. WR140][]{Williams:1990, Zhekov:2000, Pollock:2005,
     Pittard:2006}, the WCR will reside in the wind acceleration
   region of the O star's wind throughout the orbit. Consequently,
   lower preshock velocities will increase the importance of radiative
   cooling in the postshock gas, affecting the stability of the
   WCR. The interplay between the stellar radiation fields may also
   significantly reduce the acquired preshock velocities through
   radiative inhibition \citep{Stevens:1994} and/or braking
   \citep[][see also
     \citeauthor{Owocki:1995}~\citeyear{Owocki:1995}]{Gayley:1997}. In
   fact, a stable ram pressure balance may not be established between
   the winds. WR22 - one of the most massive Wolf-Rayet stars
   currently known \citep{Rauw:1996} - may play host to these
   interesting phenomena which have the potential to significantly
   affect the observed X-ray emission.

   The influence of orbital motion on the circumbinary medium may also
   affect the observed emission. The shape of the wind-wind collision
   region (WCR) between the stars is largely dependent on the ram
   pressure of the stellar winds. However, orbital motion introduces
   curvature to the WCR away from the stars \citep{Walder:1999,
     Folini:2000, Walder:2003, Lemaster:2007, Parkin:2008,
     Okazaki:2008, Parkin:2009, Pittard:2009, Parkin:2011b,
     vanMarle:2011}. This results in the spiral-like structure seen in
   the so-called ``pinwheel'' nebula \citep[e.g.][]{Monnier:1999,
     Tuthill:2006, Tuthill:2008}. As such, the resulting highly
   asymmetric gas distribution introduces a viewing angle dependence
   to the emergent X-ray spectrum.

   In this paper the wind-wind collision in WR22 is investigated using
   three dimensional adaptive-mesh refinement (AMR) simulations aimed
   at establishing the importance of the wind acceleration regions,
   radiative cooling, and orbital motion on the dynamics and resulting
   X-ray emission. The remainder of this paper is structured as
   follows: the hydrodynamical and X-ray calculations are described in
   \S~\ref{sec:model}. The simulation dynamics, resulting X-ray
   emission, and suggested revisions to model parameters are given in
   \S~\ref{sec:results}. A discussion is given in
   \S~\ref{sec:discussion}, followed by conclusions in
   \S~\ref{sec:conclusions}.

   \begin{table}
     \begin{center}
       \caption[]{Adopted system parameters for WR~22. References are
         as follows: 1 = \cite{Rauw:1996}, 2 =
         \cite{Gosset:2009}, 3 = \cite{Bohlin:1978}, 4 =
         \cite{Diplas:1994}.} \label{tab:system_parameters}
       \begin{tabular}{lll}
         \hline
         Parameter & Value & Reference \\
         \hline
         Orbital period (d) & 80.325 & 1 \\
         $a$ (au) & 1.68 & 1 \\
         Eccentricity ($e$) & 0.559 & 1 \\
         Distance (kpc) & 2.7 & 2 \\
         ISM column ($10^{21}\rm{cm}^{-2}$) & 2.5 & 3+4 \\
         \hline
       \end{tabular}
     \end{center}
   \end{table}

   \begin{table}
     \begin{center}
       \caption[]{Stellar parameters used to calculate the line
         driving of the stellar winds. $k$ and $\alpha$ are the
         \cite{Castor:1975} line driving parameters. References are as
         follows: 1 = \cite{vanderHucht:1981}, 2 =
         \cite{Schweickhardt:1999}, 3 = \cite{Rauw:1996}, 4 =
         \cite{Rauw:1997}, 5 = \cite{Hamann:2006}, 6 =
         \cite{Martins:2005}, 7 = \cite{Vink:2001}, 8 =
         \cite{Crowther:1995b}.} \label{tab:stellar_parameters}
        \begin{tabular}{lllll}
         \hline
          & \multicolumn{2}{c}{WR} & \multicolumn{2}{c}{O star} \\
         \hline
         Parameter & Value & Reference & Value & Reference \\
         \hline
         Spectral Type & WN7 & 1 & O9V & 2  \\
         $M$ (M$_{\odot}$) & 72 & 3 & 25.7 & 3 \\ 
         $R_{\ast}$ (R$_{\odot}$) & 20 & 4 & 11 & 4 \\
         $T_{\rm{eff}}$ (K)& 44700 & 5 & 33000 & 6 \\
         $L_{\ast} $ ($10^{6} {\rm L_{\odot}}$) & 1.44 &  $-$ & 0.13 & $-$ \\
         $k$ & 0.42 & $-$ & 0.30 & $-$ \\
         $\alpha$ & 0.47 & $-$ & 0.52 & $-$ \\
         $\dot{M}\;({\rm M_{\odot}~yr^{-1}})$ & $1.6\times10^{-5}$ & 4 & $2.8\times10^{-7}$ & 7 \\
         $v_{\infty}\;({\rm km~s^{-1}})$ & 1785 & 8 & 2100 & 7 \\
         \hline
       \end{tabular}
    \end{center}
   \end{table}


\section{The model}
\label{sec:model}

\subsection{Hydrodynamic modelling}
\label{subsec:hydromodel}

To model the wind-wind collision we numerically solve the
time-dependent equations of Eulerian hydrodynamics in 3D Cartesian
coordinates. The relevant equations for mass, momentum, and energy
conservation are:
\begin{eqnarray}
\frac{\partial\rho}{\partial t} + \nabla \cdot \rho {\bf v} &  =  & 0, \\
\frac{\partial\rho{\bf v}}{\partial t} + \nabla\cdot\rho{\bf vv} + \nabla P & = & \rho{\bf f},\\
\frac{\partial\rho E}{\partial t} + \nabla\cdot[(\rho E + P){\bf v}] & =& \left(\frac{\rho}{m_{\rm H}}\right)^{2}\Lambda(T) + \rho {\bf f}\cdot {\bf v}.
\end{eqnarray}

\noindent Here $E = \epsilon + \frac{1}{2}|{\bf v}|^{2}$, is the total
gas energy, $\epsilon$ is the specific internal energy, ${\bf v}$ is
the gas velocity, $\rho$ is the mass density, $P$ is the pressure, $T$
is the temperature, and $m_{\rm H}$ is the mass of hydrogen. We use
the ideal gas equation of state, $P = (\gamma - 1)\rho\epsilon$,
where the adiabatic index $\gamma = 5/3$.

\begin{figure}
  \begin{center}
    \begin{tabular}{c}
      \resizebox{80mm}{!}{\includegraphics{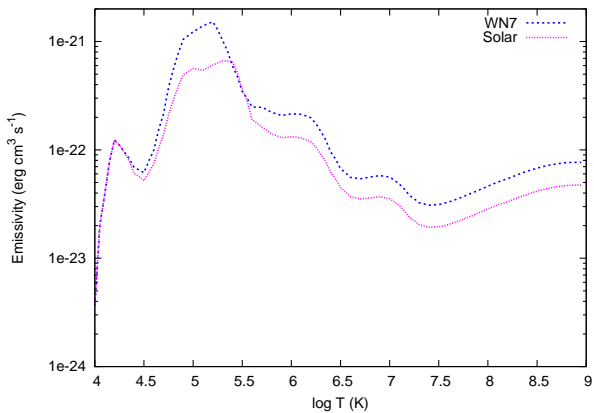}} \\
    \end{tabular}
    \caption{Cooling curves calculated for WN7 and solar abundances
      (\S~\ref{subsec:hydromodel}).}
    \label{fig:cooling_curves}
  \end{center}
\end{figure}

The radiative cooling term, $\Lambda(T)$, is calculated from the
\textsc{APEC} thermal plasma code \citep{Smith_APEC:2001} distributed
in \textsc{XSPEC} (v12.5.1). The temperature of the unshocked winds is
assumed to be maintained at $\approx 10^{4}\;$K via photoionization
heating by the stars. For the WR star we adopt the abundances of
\cite{Crowther:1995a} and \cite{Gosset:2009} \citep[see
  also][]{Hamann:1991}, and for the O star solar abundances are
assumed \citep{Anders:1989}. The cooling curves are shown in
Fig.~\ref{fig:cooling_curves}.

The body force per unit mass ${\bf f}$ acting on each hydrodynamic
cell is the vector summation of gravitational forces from each star,
and continuum and line driving forces from the stellar radiation
fields. The calculation of the line force has been described by
\cite{Pittard:2009} and \cite{Parkin:2011b}, and we refer the reader
to these works for further details \citep[see also - ][]{Cranmer:1995,
  Gayley:1997}. In brief, the numerical scheme incorporates the
\cite*{Castor:1975} \citep[hereafter ][]{Castor:1975} formalism for
line driving by evaluating the local Sobolev optical depth $\tau=
\sigma_{\rm e}v_{\rm th}\rho[\hat{\bf n}\cdot \nabla (\hat{\bf n}\cdot
  {\bf v})]^{-1}$ \citep{Sobolev:1960} and then calculating the vector
radiative force per unit mass
\begin{equation}
{\bf g}_{\rm rad} = \frac{\sigma_{e}^{1-\alpha}k}{c} \oint I(\hat{\bf n})\left[\frac{\hat{\bf n}\cdot \nabla (\hat{\bf n}\cdot {\bf v})}{\rho v_{\rm th}}\right]^{\alpha}\hat{\bf n}d{\bf \Omega},
\end{equation}
\noindent where $\alpha$ and $k$ are the standard \cite{Castor:1975}
parameters, $\sigma_{\rm e}$ is the specific electron opacity due to
Thomson scattering, and $v_{\rm th}$ is a fiducial thermal velocity
calculated for hydrogen. A gaussian integration is performed to
correct the line force for the finite size of the stellar disk
\citep[][]{Castor:1974, Pauldrach:1986}. The line driving is set to
zero in cells with temperatures above $10^{6}\;$K, since this plasma
is mostly ionized.

We note that the \cite{Castor:1975} formalism was originally developed
for O star winds, and as such, certain aspects of WR wind acceleration
are not captured. For instance, the presence of two acceleration
regions, one in the optically thick inner wind and a second in the
outer wind, and a shallower velocity profile than predicted by
\cite{Castor:1975} \citep[][]{Hillier:1999,
  Grafener:2005}. Consequently, the \cite{Castor:1975} formalism will
underestimate the column densities for lines of sight passing through
the inner WR wind ($R< 100~R_{\ast}$). However, the wind-wind
collision generally occurs close to the O star and, therefore, the
details of the WR wind acceleration should not significantly affect
the WCR dynamics. Importantly, the \cite{Castor:1975} formalism
provides a means for incorporating the wind acceleration regions.

\subsection{The hydrodynamic code}
\label{subsec:hydrocode}

The hydrodynamic equations are solved using v3.1.1 of the
\textsc{FLASH} code \citep{Fryxell:2000, Dubey:2009}. This code uses
the piecewise-parabolic method of \cite{Colella:1984} to solve the
hydrodynamic equations and operates with a block-structured AMR grid
\citep[e.g.][]{Berger:1984} using the \textsc{PARAMESH} package
\citep{MacNeice:2000} under the message-passing interface (MPI)
architecture. The simulation domain extends from $x = y =
\pm1.2\times10^{14}\;$cm and $z = (0 - 1.2\times10^{14})\;$cm - a
symmetry about the $xy$-plane is used. The grid is initialized with $x
\times y \times z = 16 \times 16 \times 8$ cubic blocks each
containing $8^{3}$ cells. We allow for 5 nested levels of refinement,
which results in an effective resolution on the finest grid level of
$x \times y \times z = 4096 \times 4096 \times 2048\;$cells.  The
refinement of the grid depends on a second-derivative error check
\citep{Fryxell:2000} on $\rho$ and the requirement of an effective
number of cells between the stars to accurately describe the WCR
dynamics \citep{Parkin:2011b}. Customized units have been implemented
into the \textsc{FLASH} code for radiative driving, gravity, orbital
motion, and radiative cooling for optically-thin plasma \citep[using
  the explicit method described in][]{Strickland:1995}. We include an
advected scalar variable to track the stellar winds.

The stellar winds are initiated in the instantaneously accelerated and
radiatively driven stellar winds simulations in two slightly different
ways and we refer the reader to \cite{Pittard:2009} and
\cite{Parkin:2011b} for further details. In short, the instantaneously
accelerated winds are initiated into a radius which adapts to the
simulation resolution (which is dependent on the separation of the
stars), whereas the radiatively-driven winds are initiated into a
radius of $\sim 1.15\;{\rm R_{\ast}}$ around the stars at all orbital
phases.

The orbital motion of the stars is calculated in the centre of mass
frame. When the stars are at apastron the WR and O stars are situated
on the positive and negative $x-$axis, respectively. The motion of the
stars proceeds in an anti-clockwise direction.

\begin{figure*}[!]
  \begin{center}
    \begin{tabular}{cc}
\resizebox{80mm}{!}{\includegraphics{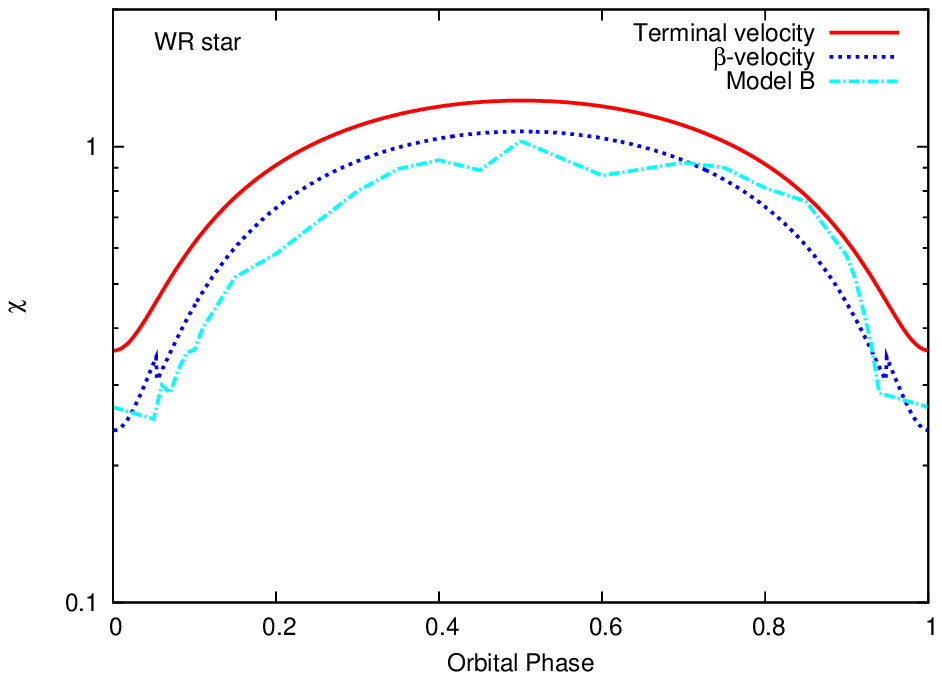}} &
\resizebox{80mm}{!}{\includegraphics{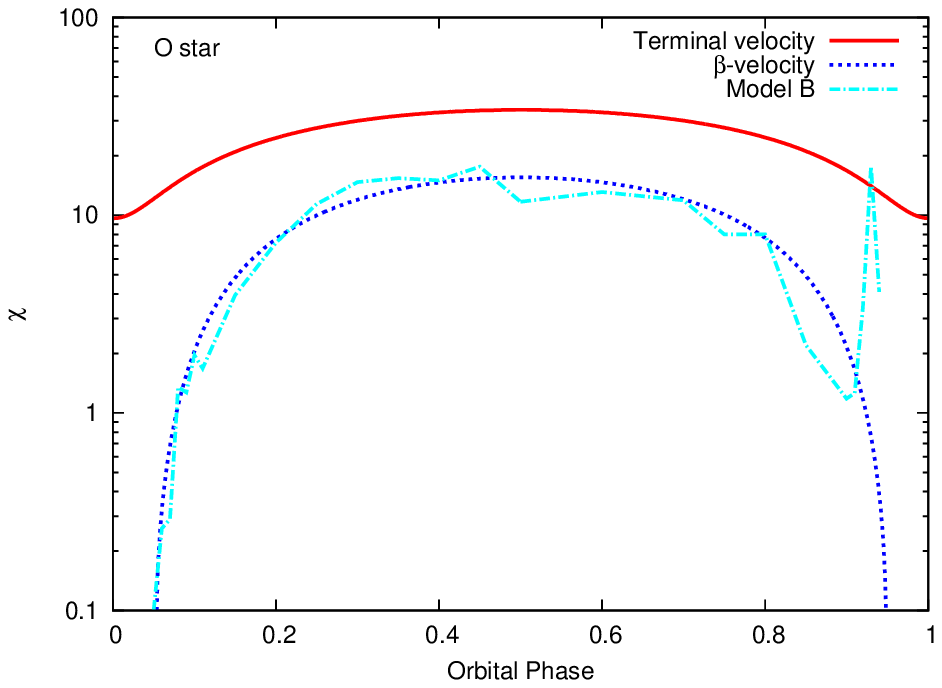}}  \\
    \end{tabular}
    \caption{Cooling parameter, $\chi$, calculated for the WR's wind
      (left panel) and O star's wind (right panel), respectively. Note
      the difference in scale between the plots.}
    \label{fig:chi}
  \end{center}
\end{figure*}

\subsection{X-ray emission}
\label{subsec:xray_emission}

Synthetic X-ray spectra and lightcurves are produced by solving the
equation of radiative transfer through the simulation domain using
adaptive image ray-tracing (AIR). An initially low resolution image
equivalent to that of the base hydrodynamic grid
(i.e. $128\times128\;$pixels) is constructed. The image is then
scanned using a second-derivate truncation error check on the
intrinsic 0.5-10~keV X-ray flux and sufficiently prominent pixels
\citep[$\xi_{\rm crit} = 0.6$, where $\xi_{\rm crit}$ is the critical
  truncation error above which pixels are marked for refinement - see
][]{Parkin:2011} are refined. The process of ray-tracing and refining
pixels is repeated until features of interest in the image have been
captured to an effective resolution equivalent to that of the
simulation (i.e. $4096\times 4096\;$pixels). For details of the AIR
technique we refer the reader to \cite{Parkin:2011}.

To calculate the X-ray emission from the simulation we use
emissivities for optically thin gas in collisional ionization
equilibrium obtained from look-up tables calculated from the
\textsc{APEC} plasma code \citep{Smith_APEC:2001} containing 200
logarithmically spaced energy bins in the range 0.1-10 keV, and 101
logarithmically spaced temperature bins in the range
$10^{4}-10^{9}\;$K. When calculating the emergent flux we use energy
dependent opacities calculated with version $c08.00$ of Cloudy
\citep[][see also
  \citeauthor{Ferland:1998}~\citeyear{Ferland:1998}]{Ferland:2000}. The
advected scalar is used to separate the X-ray emission contributions
from each wind.

\section{Results}
\label{sec:results}

For the purposes of our investigation we have performed two
simulations: one in which the stellar winds are assumed to be
instantaneously accelerated at the surface of the star (model A), and
another in which the winds are radiatively driven (model B). Our
adopted orbital and stellar parameters are noted in
Tables~\ref{tab:system_parameters} and
\ref{tab:stellar_parameters}. To form a basis for the analysis of the
hydrodynamical simulations, we first estimate the influence of the
wind acceleration regions on the WCR dynamics. The dynamics and X-ray
emission calculations from models A and B are then presented,
following which we consider some potential revisions to the adopted
parameters for WR22.

\subsection{Estimating the impact of wind acceleration}
\label{subsec:estimates}

In model A the stellar winds collide at their terminal velocity
throughout the orbit, albeit with a small additional velocity
component due to contraction/expansion of the binary separation. The
wind momentum ratio $\eta =\dot{M}_{\rm O}v_{\rm O}/\dot{M}_{\rm
  WR}v_{\rm WR} = 0.02$, and the relative distance of the WCR from the
stars (along the line of centres), stays roughly constant\footnote{The
  thermal pressure of postshock gas offsets the position of the
  stagnation point from that attained by the ram pressure of the winds
  alone \citep[e.g.][]{Kenny:2005, Gayley:2009}. Due to the variation
  in radiative cooling for the postshock gas through the orbit the
  offset due to postshock thermal pressure also
  varies.}. Consequently, a stable ram pressure balance exists between
the winds at all orbital phases. The WCR dynamics are largely dictated
by the importance of radiative cooling \citep[see
e.g.][]{Stevens:1992, Parkin_Pittard:2010} which can be quantified
using the cooling parameter $\chi$ \citep{Stevens:1992}\footnote{The
  cooling parameter $\chi = t_{\rm cool}/t_{\rm flow} = v_{8}^4
  d_{12}/ \dot{M}_{-7}$, where $t_{\rm cool}$ is the cooling time,
  $t_{\rm flow}$ is the flow time, $v_{8}$ is the preshock gas
  velocity (in $10^{8}\;$cm~s$^{-1}$), $d_{12}$ is a characteristic
  distance (in $10^{12}\;$cm) taken here to be the stagnation point
  radius, and $\dot{M}_{-7}$ is the mass-loss rate (in $10^{-7}\;{\rm
    M_{\odot}~yr^{-1}}$). $\chi$ is the ratio of the cooling time to
  the flow time; $\chi > 1$ indicates quasi-adiabatic gas, whereas
  $\chi\ltsimm 1$ indicates that postshock gas cools rapidly. Note
  that in our calculation of $\chi_{\rm WR}$ we have accounted for the
  emissivity of WN7 abundance gas at $10^{7}\;$K being roughly a
  factor of 1.7 higher than that of solar abundance gas (see
  Fig.~\ref{fig:cooling_curves}), hence our values of $\chi_{\rm WR}$
  are lower than those quoted by \cite{Gosset:2009}.}. The postshock
WR's wind will cool rapidly at all orbital phases ($\chi_{\rm
  WR}\simeq 0.35 - 1.3$ - Fig.~\ref{fig:chi}), whereas the postshock O
star wind will be quasi-adiabatic throughout the orbit ($\chi_{\rm
  O}\simeq 10-30$). The velocity and density shear across the contact
discontinuity (CD) will cause Kelvin-Helmholtz (KH) instabilities to
develop. As the postshock WR wind cools to form a thin dense layer,
ripples introduced by KH and Rayleigh-Taylor (RT) instabilities will
seed the growth of thin-shell instabilities
\citep{Vishniac:1983}. However, the hot, quasi-adiabatic postshock O
star wind will act as a cushion and prevent the evolution of the
bending modes to non-linear thin-shell instabilities \citep[NTSIs
-][]{Vishniac:1994}.

\begin{figure}[!]
  \begin{center}
    \begin{tabular}{c}
\resizebox{80mm}{!}{\includegraphics{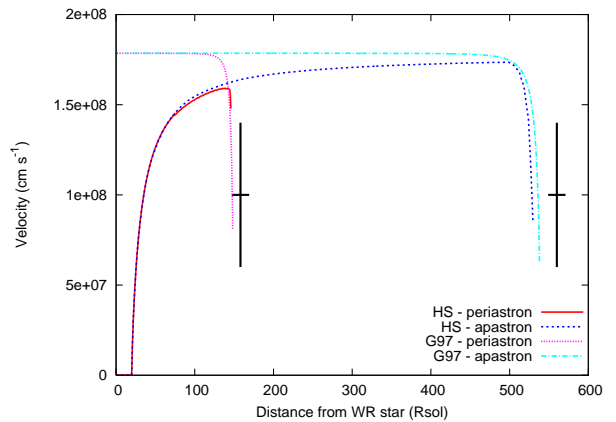}} \\
    \end{tabular}
    \caption{The WR's wind velocity along the line of centres
      calculated using hydrodynamic simulations with radiatively
      driven stellar winds (HS), and using the analytical radiative
      braking prescription of \cite{Gayley:1997} (G97). The crosses
      indicate the position of the O star at periastron ($\sim
      158\;{\rm R_{\odot}}$) and apastron ($\sim560\;{\rm
        R_{\odot}}$), and the cross width corresponds to the diameter
      of the O star. Note that orbital motion is not included in these
      calculations.}
    \label{fig:radiative_braking}
  \end{center}
\end{figure}

The inclusion of the wind acceleration regions introduces a number of
important differences to the WCR dynamics. Firstly, the separation of
the stars is sufficiently small that the winds do not reach their
terminal velocities before they collide. Therefore, the {\it
  effective} wind momentum ratio is a function of orbital phase as it
depends on the ram pressure balance between the accelerating
winds. Because the ram pressure of the WR wind dominates, the WCR (if
one exists) will occur closer to the O star than in the
instantaneously accelerated winds case. Hence, at phases close to
periastron the WCR will reside deep in the acceleration region of the
O star's wind, and the acquired preshock velocities will be well below
the terminal wind velocity. The impact of wind acceleration on the
importance of cooling in the postshock gas can be estimated if one
approximates the winds as $\beta$-velocity laws
(i.e. $v(r)=v_{\infty}(1 - R_{\ast}/r)^{\beta}\;$with
$\beta=1$)\footnote{Observational and theoretical studies suggest that
  $\beta > 1$ is appropriate for WR winds
  \citep[e.g.][]{Koenigsberger:1990, Auer:1994, Springmann:1994,
    Gayley:1995, Hillier:1999, Grafener:2005}. However, as the
  numerical scheme for the radiatively-driven winds is based on
  \cite{Castor:1975}, adopting $\beta=1$ allows predictions to be made
  for the results of model B.}  and solves for the ram pressure
balance between the winds to determine effective values for $\chi$. In
this case, the postshock O star's wind remains quasi-adiabatic for the
majority of the orbit ($0.08\ltsimm \phi \ltsimm 0.92$) but cooling
becomes important at phases close to periastron
(Fig.~\ref{fig:chi}). Furthermore, at $\phi\simeq0.95$ a ram pressure
balance is not attained between the winds and the WR wind now collides
against the O star - a corresponding step in $\chi_{\rm WR}$ occurs
indicating the sudden increase in the stagnation point
radius. \cite{Gayley:1997} note that for WR22 a WCR may be established
due to the radiative braking of the WR wind by the O star. Evaluating
the equations of \cite{Gayley:1997}, we find that radiative braking
should occur at periastron and apastron
(Fig.~\ref{fig:radiative_braking} - although at apastron the WR wind
has reached the WCR before the onset of braking). For comparison,
hydrodynamic simulations with radiatively driven winds, and
separations corresponding to apastron and periastron, have been
performed\footnote{In these simulations the O star radiation field is
  included but its wind is not, hence no wind-wind collision shocks
  occur.} (using the code described in \S~\ref{sec:model}). Although
the WR's wind is not as effectively halted before it reaches the O
star in the hydrodynamic simulation at periastron separation, the
calculations agree in that some braking should occur. Note that in
these calculations we have used the O star's \cite{Castor:1975}
parameters ($k_{\rm O}= 0.30$ and $\alpha_{\rm O}= 0.52$) to assess
the radiative braking of the WR's wind. If the WR's \cite{Castor:1975}
parameters were used (i.e. $k_{\rm O}= 0.42$ and $\alpha_{\rm O}=
0.47$) radiative braking would be more effective.

Notwithstanding the possibility of radiative braking, the importance
of cooling for both winds will permit the growth of NTSIs close to the
WCR apex. The subsequent disruption, and possible collapse, of the WCR
will significantly affect the dynamics and resulting X-ray emission.


\subsection{Model A Dynamics}
\label{subsec:modelA}

\begin{figure*}
  \begin{center}
    \begin{tabular}{cc}
\resizebox{60mm}{!}{\includegraphics{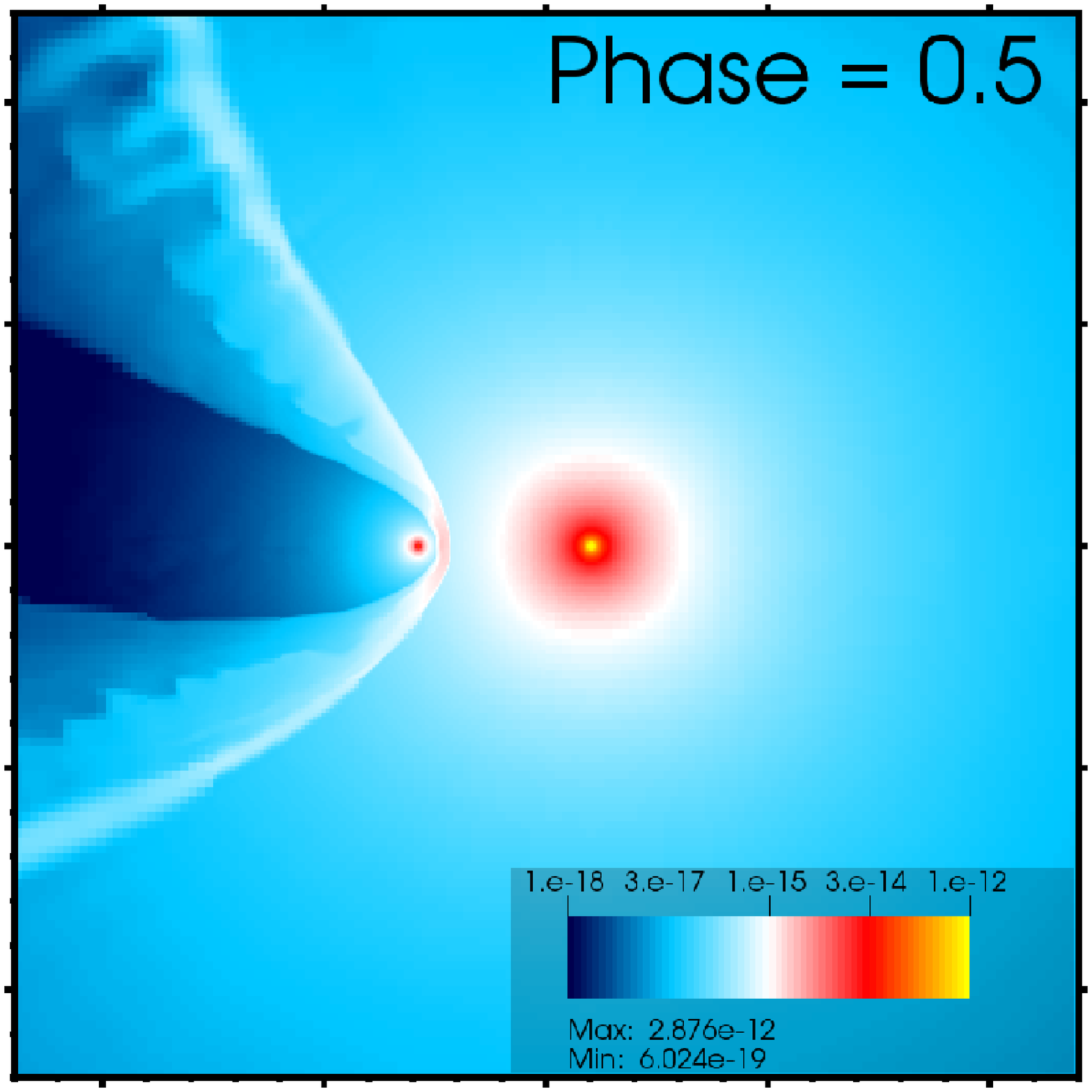}} & 
\resizebox{60mm}{!}{\includegraphics{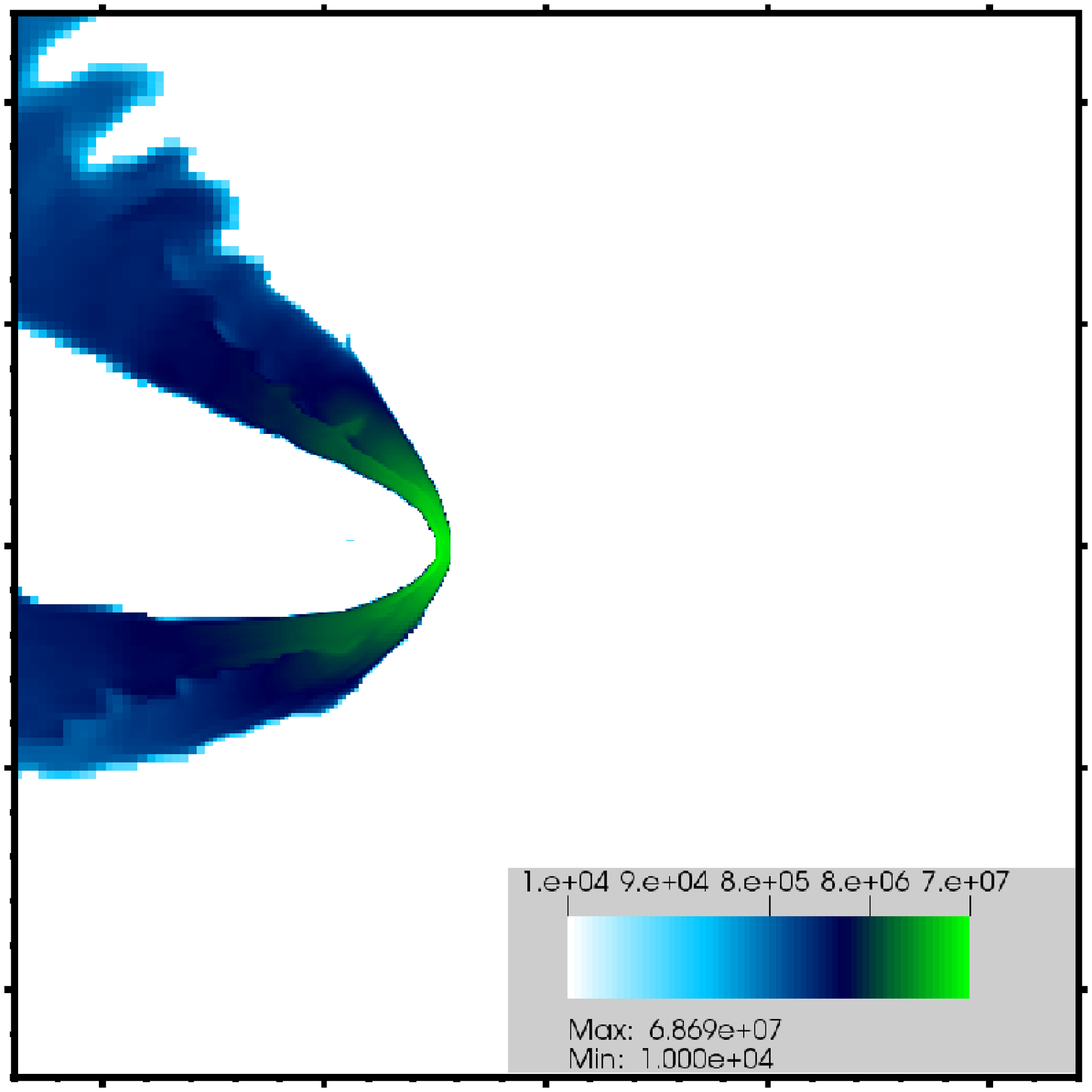}} \\
\resizebox{60mm}{!}{\includegraphics{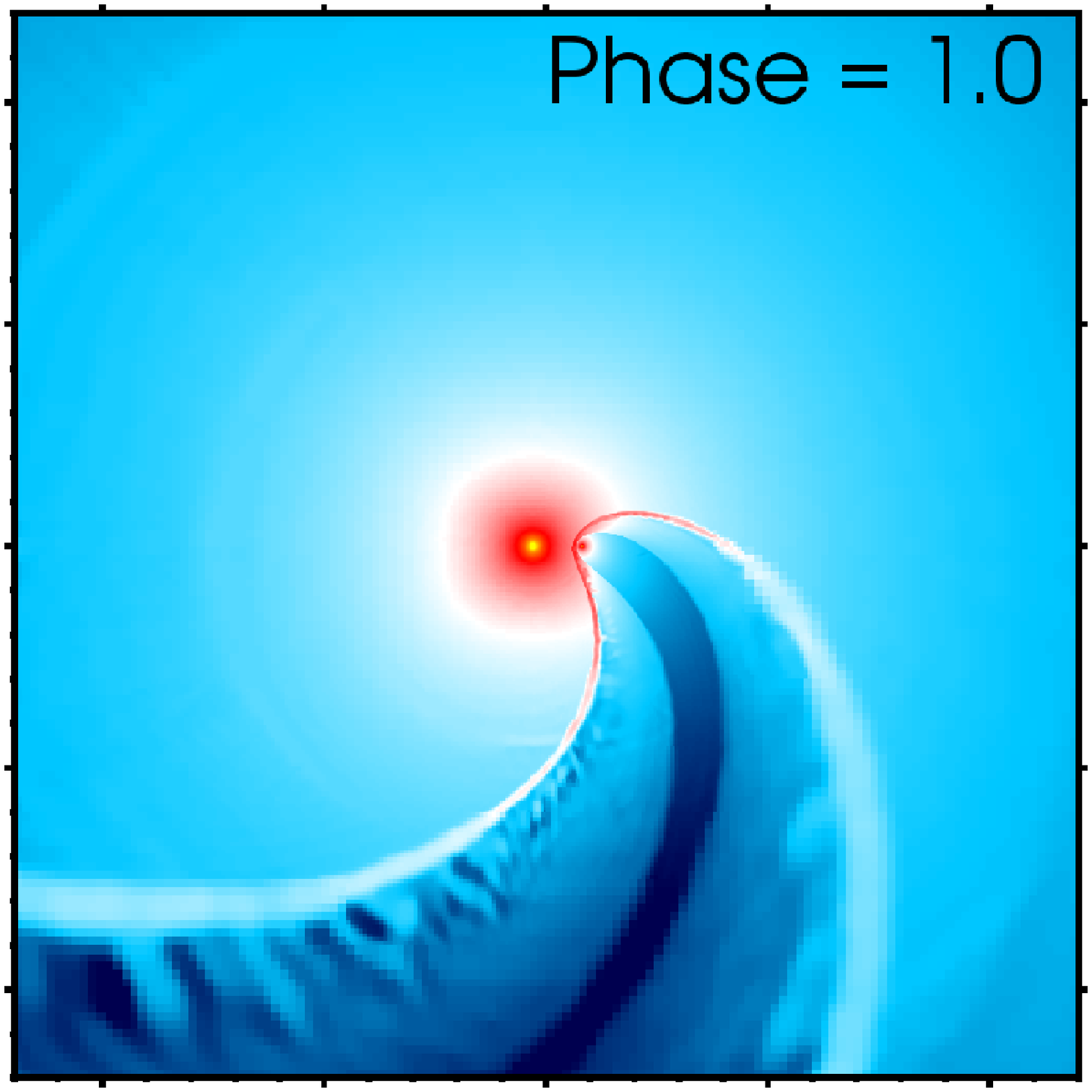}} & 
\resizebox{60mm}{!}{\includegraphics{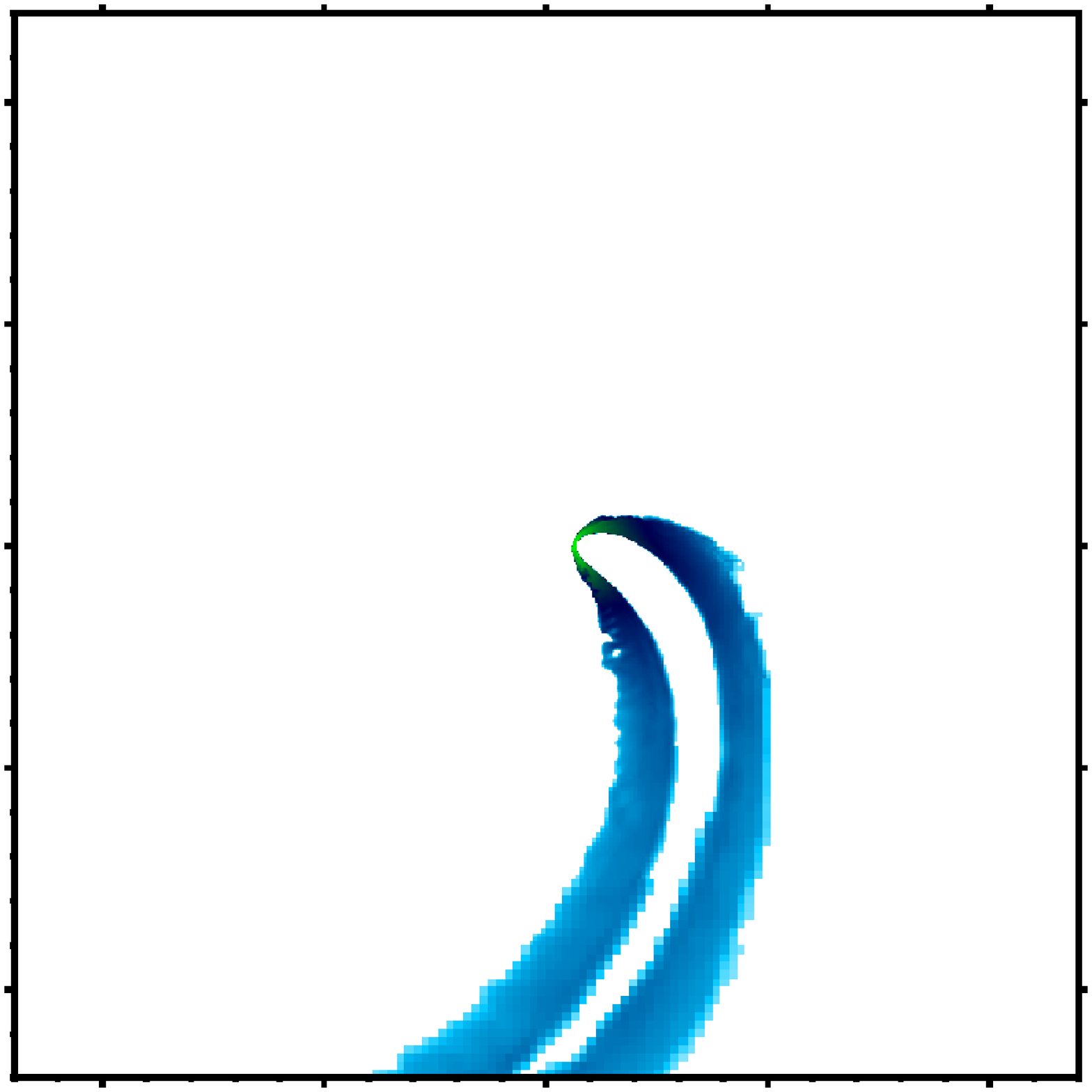}}\\
\resizebox{60mm}{!}{\includegraphics{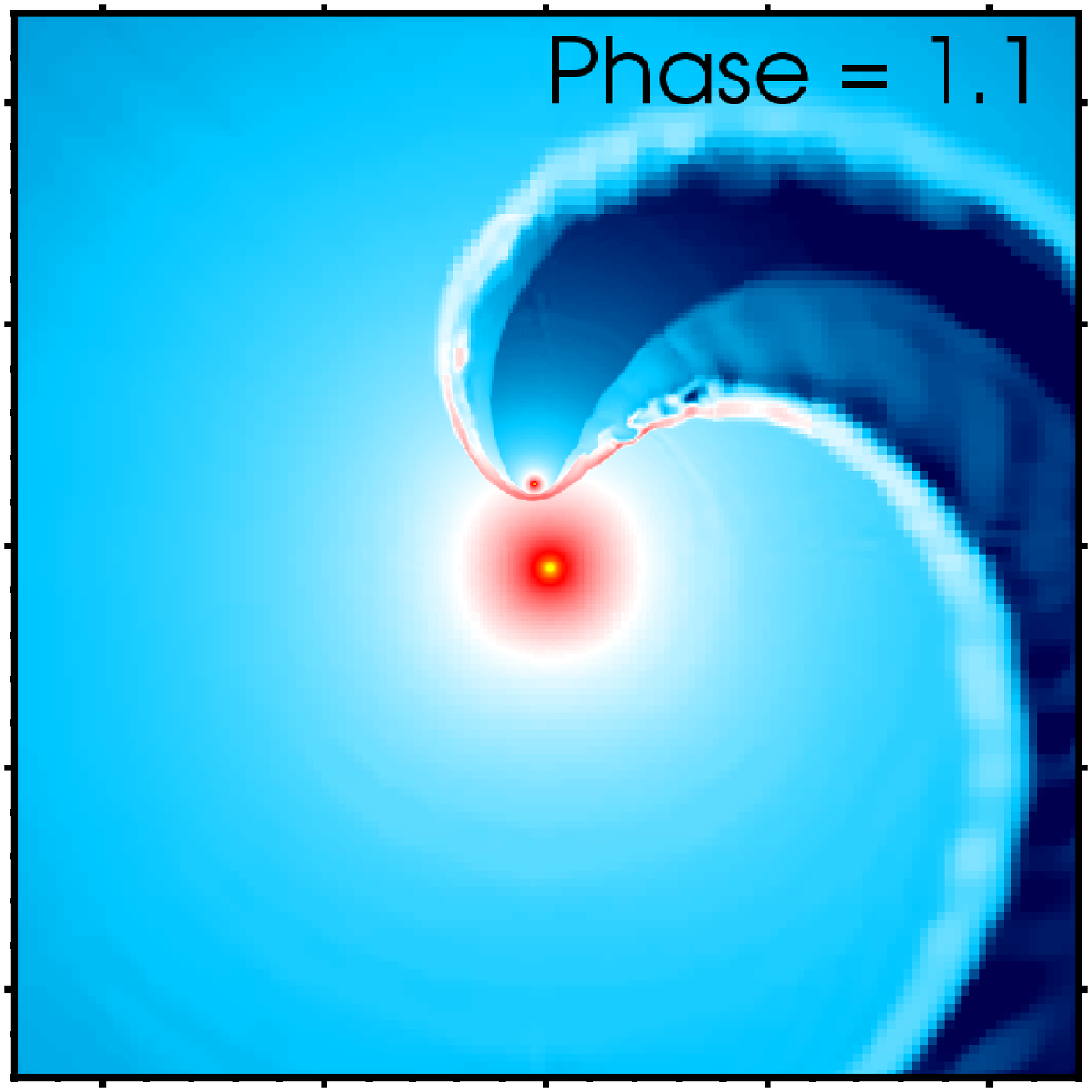}} & 
\resizebox{60mm}{!}{\includegraphics{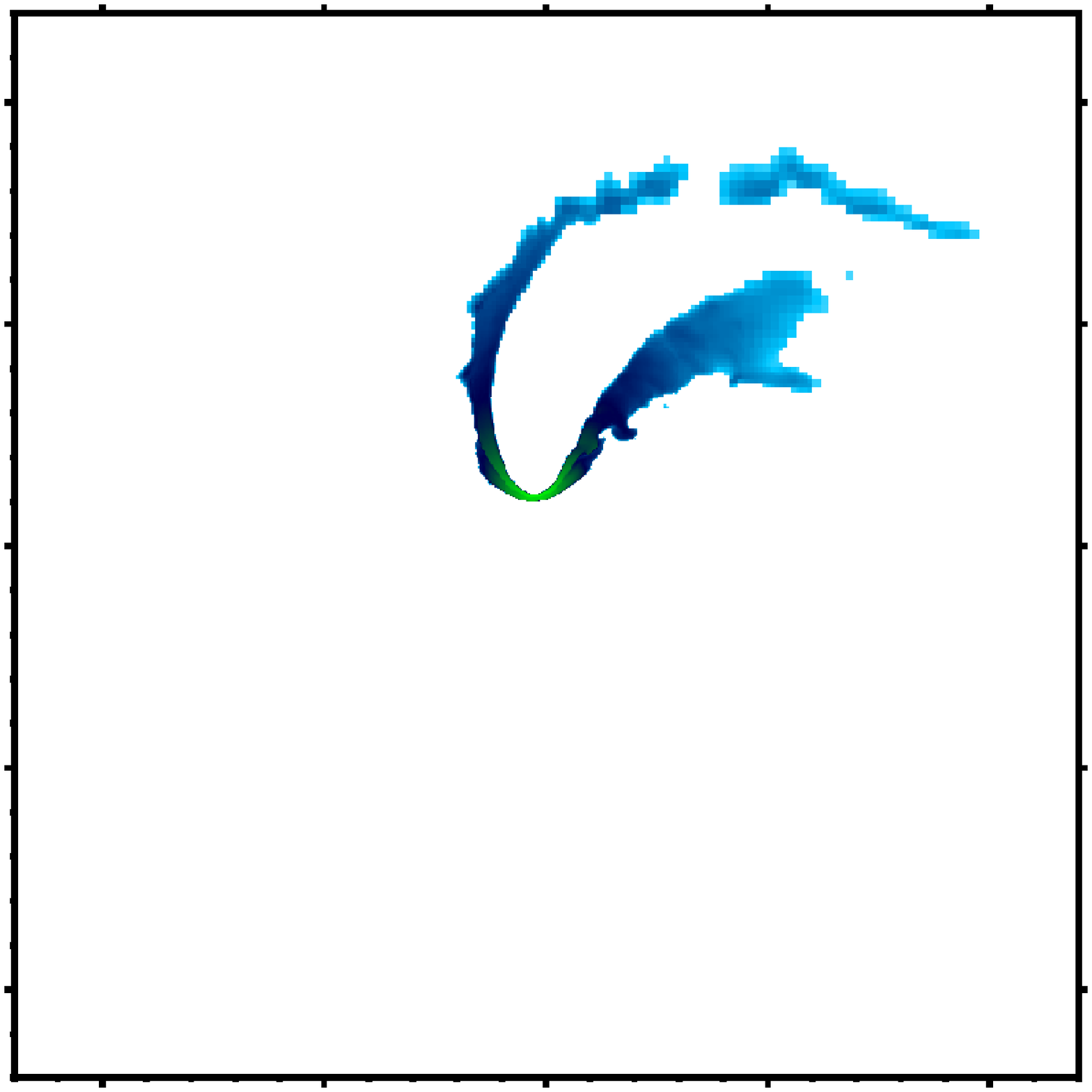}}\\
    \end{tabular}
    \caption{Snapshots of the gas density (left column) and
      temperature (right column) in the orbital ($x-y$) plane from
      model A (instantaneously accelerated winds) at $\phi =
      0.5\;$(upper panels), 1.0 (middle panels), and 1.1 (lower
      panels). The orbital motion of the stars is calculated in the
      centre of mass frame. At apastron ($\phi =0.5$) the WR star is
      to the right, and the O star is to the left, of the image
      centre. The motion of the stars proceeds in an anti-clockwise
      direction. All plots show a region of $\pm1.2\times10^{14}\;$cm
      - large axis tick marks correspond to a distance of
      $5\times10^{13}\;$cm.}
    \label{fig:vterm_images}
  \end{center}
\end{figure*}

The stronger WR wind causes the WCR to form a bow-shock which is
concave from the perspective of the O star. At apastron one sees that
the influence of orbital motion on the arms of the WCR is relatively
minor due to the low velocities of the stars compared to the wind
velocities (Fig.~\ref{fig:vterm_images}). The postshock O star wind is
quasi-adiabatic, whereas the postshock WR wind is reasonably hot close
to the apex of the WCR, but cools rapidly downstream to a temperature
of $\sim 10^{4}\;$K. A comparison of the density and temperature
snapshots in Fig.~\ref{fig:vterm_images} highlights the region of the
WCR composed of cool postshock WR wind. The velocity and density shear
across the CD causes KH instabilities to develop and more-so in the
trailing arm of the WCR due to the slightly higher shock obliquity
close to the apex. As the stars move toward periastron the influence
of orbital motion on the WCR becomes gradually more noticeable and the
degree of curvature to the WCR arms increases. The curvature of the
WCR reaches a maximum at periastron corresponding to the peak in the
orbital velocities. Interestingly, as the stars approach periastron
the postshock WR wind cools more rapidly, and the density contrast
across the CD increases. This, combined with the oblique angle at
which the WR wind collides against the trailing arm of the WCR,
results in perturbations to the CD growing to such an extent that
filamentary structure is formed by compressions/rarefactions. We note
that a similar feature was observed in simulations of $\eta~$Carinae
by \cite{Parkin:2011b}, but was not found in simulations of shorter
period OB star systems by \cite{Pittard:2009}, or in the circular
orbit simulations presented by \cite{vanMarle:2011}, suggesting that
it may only be active when the orbital eccentricity is relatively
high, and hence the trailing arm of the WCR is considerably curved
around periastron. Although some of the postshock O star's wind
remains quasi-adiabatic as it flows off the grid, the region close to
the contact discontinuity cools more effectively. This is in part due
to the mixing introduced by KH instabilities, but also results from
some numerical heat conduction between the hot and cold postshock gas
\citep{Parkin_Pittard:2010}.

\begin{figure*}
  \begin{center}
    \begin{tabular}{cc}
\resizebox{60mm}{!}{\includegraphics{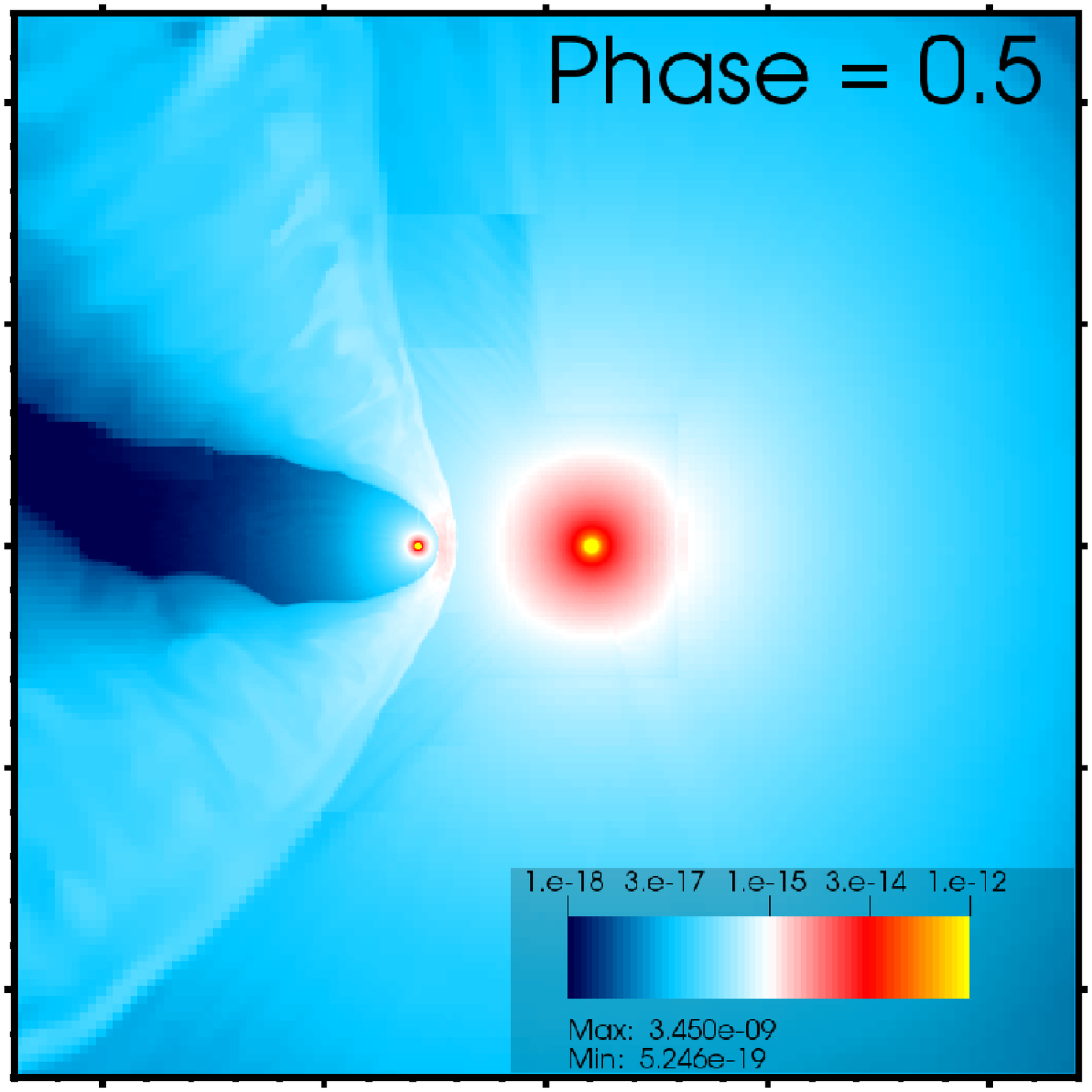}} & 
\resizebox{60mm}{!}{\includegraphics{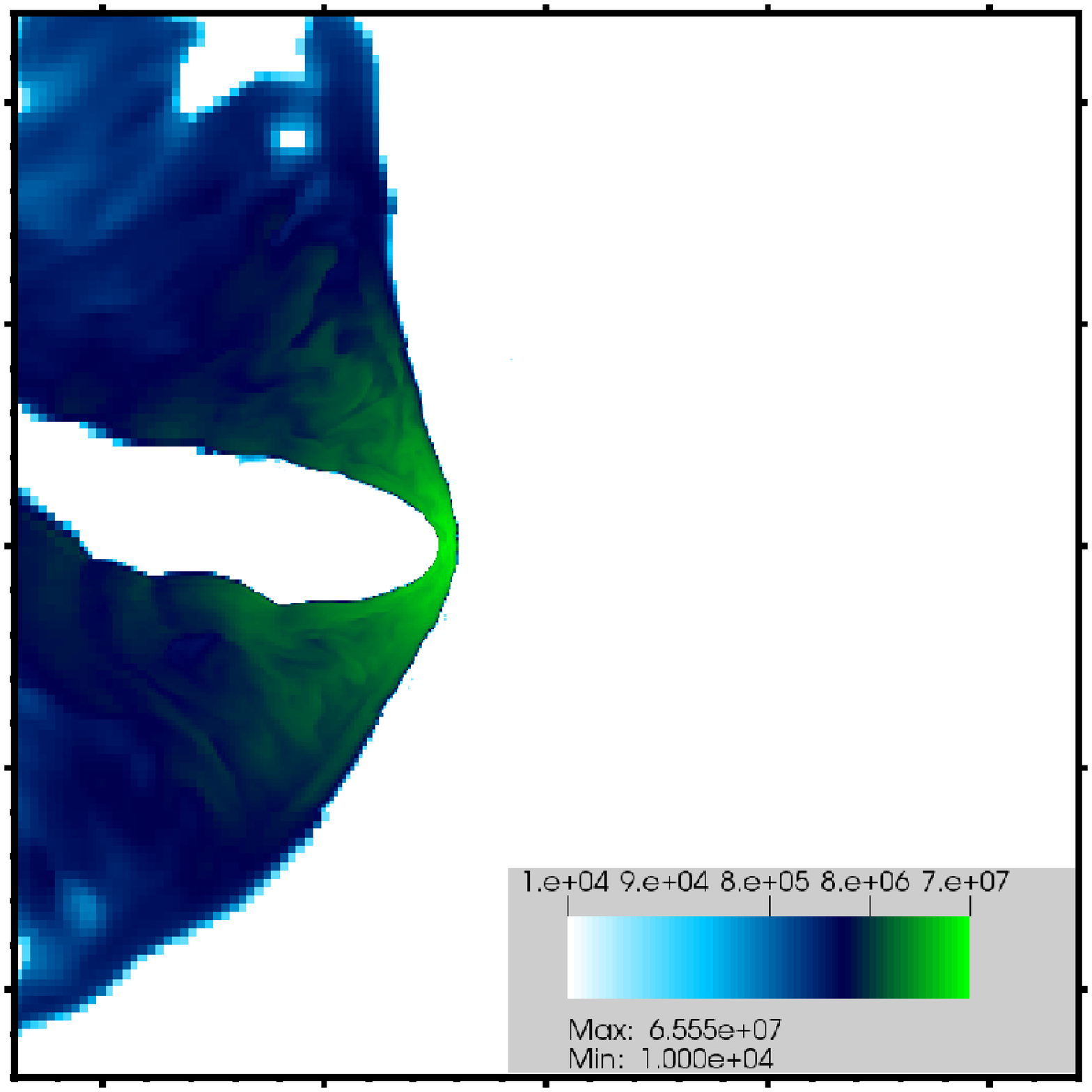}} \\
\resizebox{60mm}{!}{\includegraphics{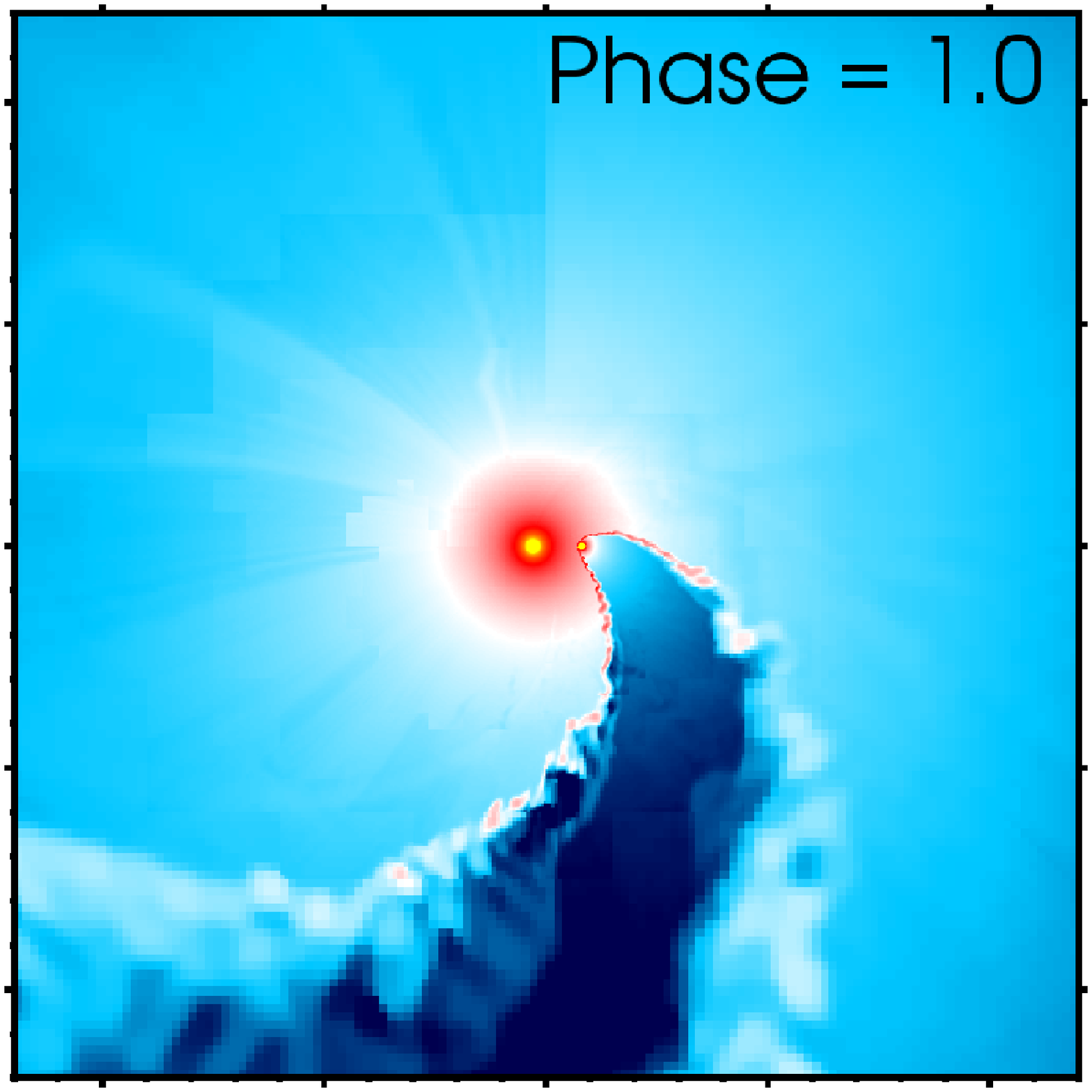}} & 
\resizebox{60mm}{!}{\includegraphics{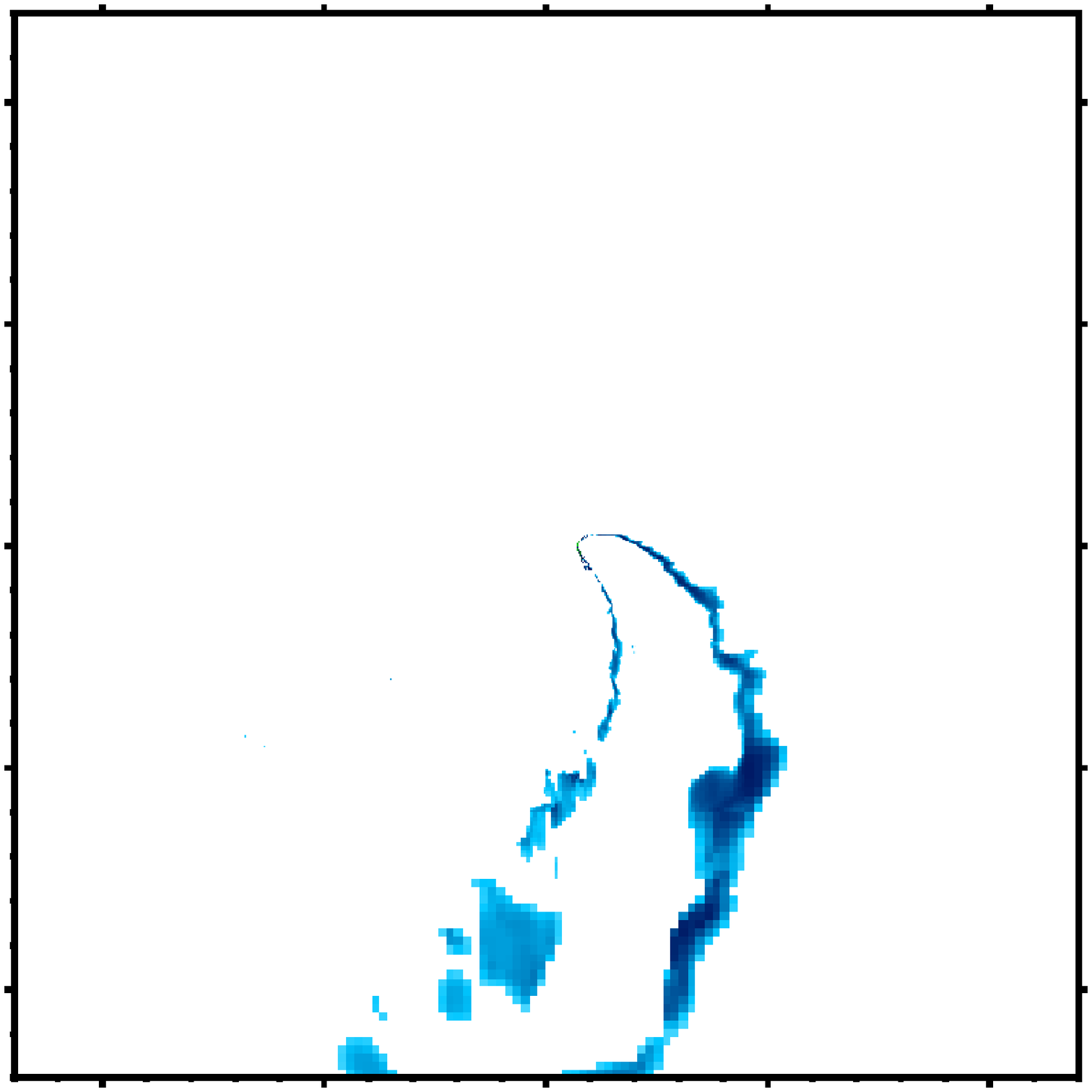}}\\
\resizebox{60mm}{!}{\includegraphics{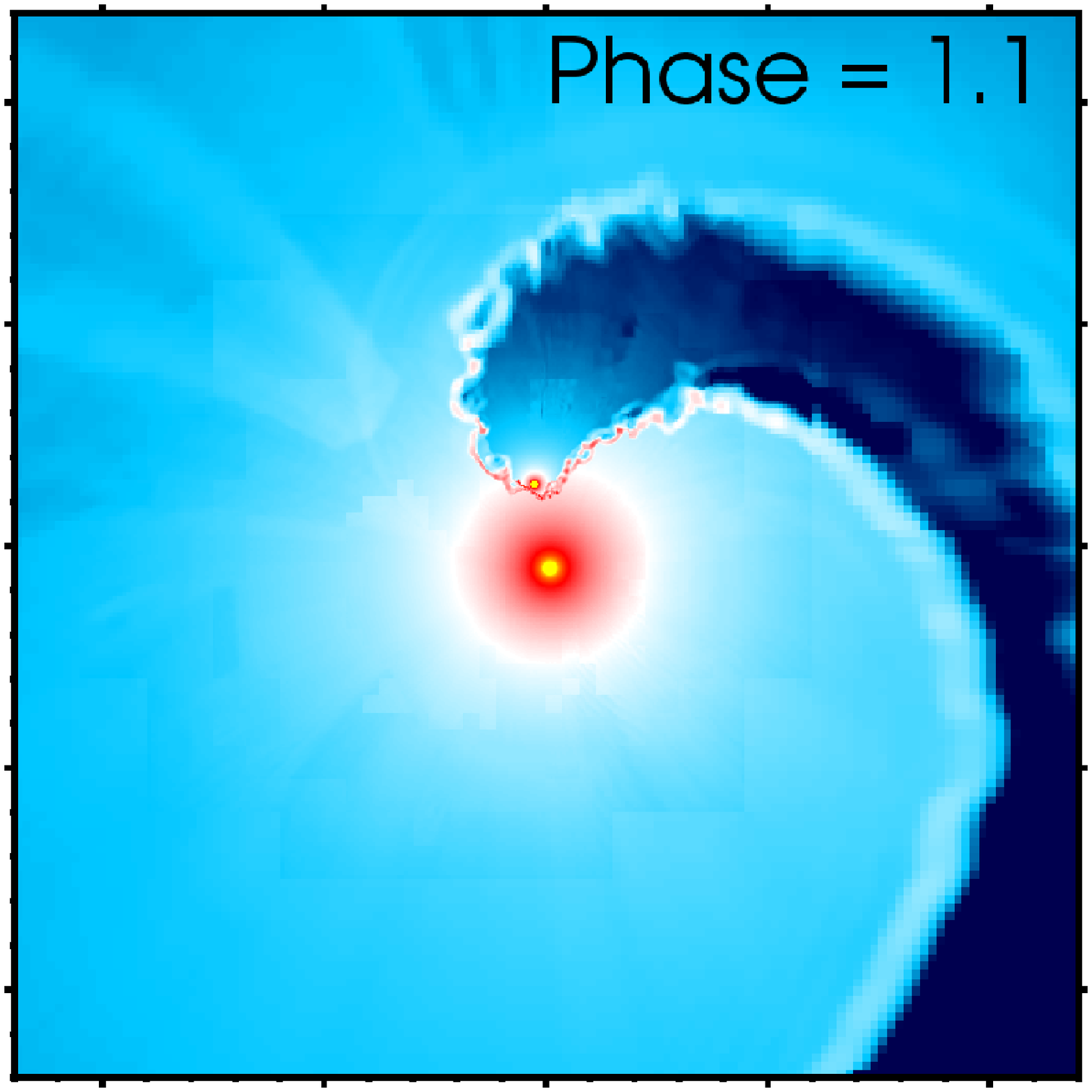}} & 
\resizebox{60mm}{!}{\includegraphics{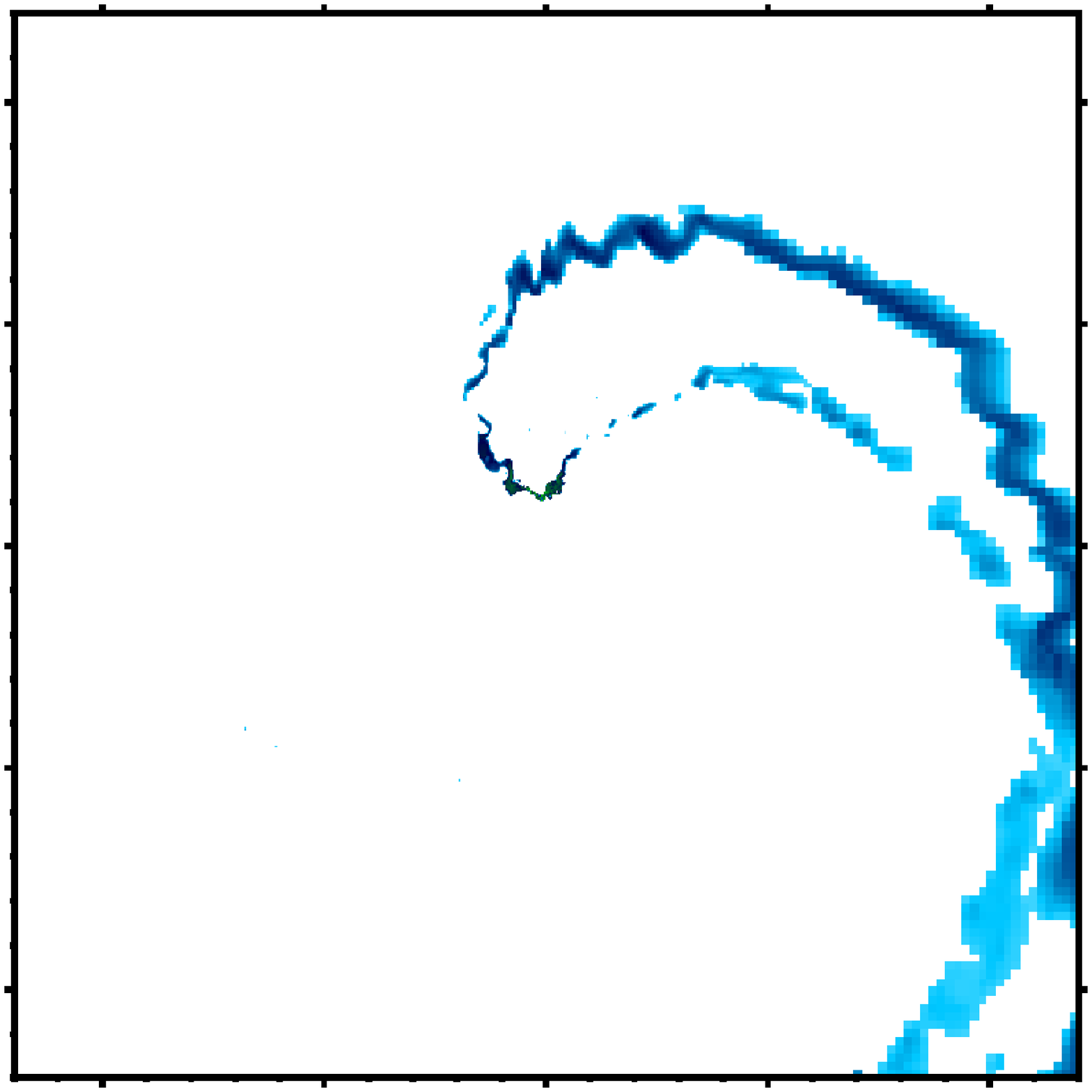}}\\
    \end{tabular}
    \caption{Same as Fig.~\ref{fig:vterm_images} except model B
      (radiatively driven winds) is shown.}
    \label{fig:driven_images}
  \end{center}
\end{figure*}

Following periastron, the WCR is wound around the stars by the
influence of their orbital motion, and the cavity in the wake of the O
star resembles an archimedian spiral ($\phi=1.1$ in
Fig.~\ref{fig:vterm_images}) reminiscent of the so-called ``pinwheel''
nebula \citep[e.g.][]{Tuthill:2008}. Examining the temperature
snapshot at $\phi=1.1$ in Fig.~\ref{fig:vterm_images} shows that
postshock O star wind in the arms of the WCR cools to $T\simeq
10^{4}\;$K before exiting the grid. This gas was shocked at an earlier
phase when $\chi_{\rm O}$ was lower, however, $\chi_{\rm O}$ was
sufficiently large for the stars to have rotated in their orbits
before cooling becomes apparent in the downstream gas.

As the stars continue in their orbit the spiralling WCR gradually
drifts outwards and exits the grid - a full rotation does not fit
inside the simulation domain. As shown by \cite{Parkin:2008}, the
coils of the WCR will extend out to large distances ($\simeq
1000\;{\rm au}$). Heading towards apastron the orbital velocities
decrease and the impact of orbital motion on the WCR gradually fades,
returning once again to an approximately axisymmetric shape.

\subsection{Model B Dynamics}
\label{subsec:modelB}

\subsubsection{Large-scale phenomenon}

In model B the stellar winds are radiatively driven, and therefore
consideration is given to their acceleration regions. Similar to model
A, at apastron in model B the postshock winds are quasi-adiabatic and
the effects of orbital motion are relatively minor. However, even at
apastron the radiation fields of the stars introduce some noticeable
differences. For instance, there is a small amount of radiative
inhibition of the preshock WR wind by the O star which causes the
distance between the WR and its respective shock to be slightly
smaller in model B compared to model A
(Fig.~\ref{fig:driven_images}). Also, due to the lower off-axis
velocities, the opening angle of the shocks are larger in model B
compared to model A.

Proceeding towards periastron, substantial differences between models
A and B become apparent. This is essentially tied to the increasing
importance of cooling in the postshock gas as the WCR moves deeper
into the wind acceleration regions (\S~\ref{subsec:estimates}). For
instance, between $\phi \simeq 0.8-0.9$ there is a sharp increase in
the importance of cooling in the postshock O star's wind and
consequently thin shell instabilities become more apparent at the apex
of the WCR. The increasing level of radiative inhibition, which occurs
as the stellar separation contracts, assists the disruption of the WCR
by further reducing the O star's preshock wind
velocity. Interestingly, the cooling parameters calculated from model
B agree very well with the estimates in \S~\ref{subsec:estimates}
based on approximating the wind acceleration using $\beta-$velocity
laws (Fig.~\ref{fig:chi}). There is, however, a spike at $\phi \simeq
0.93$ corresponding to instabilities perturbing the position of the
WCR and thus the acquired preshock velocities.

As the stellar separation contracts, the WCR moves closer to the O
star and its postshock wind begins to cool effectively, forming a
thin, dense layer. When both winds cool effectively postshock,
thin-shell instabilities at the CD \citep{Vishniac:1983} are permitted
to grow in amplitude. Essentially, this occurs because the restoring
force provided by thermal pressure in the postshock gas becomes
insufficient to prevent bending modes from evolving non-linearly
\citep{Vishniac:1994}. This contrasts with model A, where the
postshock O star's wind is quasi-adiabatic throughout the orbit, and
the WCR is supported against the growth of NTSIs because the high
thermal pressure of the postshock O star wind acts like a cushion and
suppresses the growth of ripples in the CD.

The onset of effective radiative cooling for both winds in model B
also considerably changes the downstream flow in the WCR arms, making
it far more structured/clumpy than in model A (see
$\phi=1.1\;$snapshots in Figs.~\ref{fig:vterm_images} and
\ref{fig:driven_images}). During the transition from quasi-adiabatic
to radiative postshock gas the combination of cooling and compression
by KH instabilities forms clumps in the downstream flow (see
$\phi=0.90$ snapshot in Fig.~\ref{fig:collapse_images}). Moreover, the
vigorous action of the NTSI breaks-up the thin-shell of postshock gas
as it flows away from the apex of the WCR. Interestingly, although the
NTSI was found to develop in the simulations of \cite{Pittard:2009}
and \cite{vanMarle:2011}, such excessive fragmentation was not
observed, perhaps due to a lower resolution of the postshock
flow\footnote{As demonstrated by \cite{Parkin_Pittard:2010}, the
  degree of fragmentation of the thin shell increases, and the size of
  the resulting clumps decreases, with the increasing resolution of
  the postshock flow.}. Before (after) periastron the trailing
(leading) arm of the WCR is more structured than the leading
(trailing) arm due to the larger shock obliquity and thus more
effective growth of KH instabilities (see the $\phi=0.97\;$and 1.05
snapshots in Figs.~\ref{fig:collapse_images} and
\ref{fig:recovery_images}).

As the stars proceed towards apastron the continuing injection of
thermal pressure at the apex gradually restores the stability of the
downstream shocks and smoothes out the clumpy structure in WCR
arms. As the stars proceed towards apastron the postshock winds become
quasi-adiabatic again. The relatively rapid destruction of clumps in
the WCR following periastron contrasts with the simulations of shorter
period OB star systems presented by \cite{Pittard:2009}, in which
clumps formed close to periastron survived until apastron.

\subsubsection{The collapse/recovery of the wind-wind collision region}
\label{subsec:collapse}

The growth of NTSIs corrugates the WCR apex, increasing the shock
obliquity, thereby reducing the efficiency with which wind kinetic
energy is thermalized. Consequently, postshock gas temperatures are
further reduced and cooling becomes even more effective. The runaway
disruption continues as periastron is approached and the WCR apex
oscillates wildly in the vicinity of the O star until a collision
occurs at $\phi \simeq 0.95$. The preshock O star's wind, which
suffers substantial radiative inhibition at phases close to
periastron, has little room to accelerate and its ram pressure is
unable to hold back the incoming WR wind, leading to a
collapse\footnote{During a collapse of the WCR onto the O star, the
  WR's wind will enter deep into the O star's wind acceleration
  region, and may potentially collide against the O star's
  photosphere. We do not expect that this will lead to significant
  accretion and mass transfer as the WR's wind has too great a kinetic
  energy to become bound to the O star
  (c.f. \S~\ref{subsec:accretion}).} of the WCR onto the O star at
$\phi=0.96$. We note that contrary to the predictions of
Fig.~\ref{fig:radiative_braking} radiative braking does not prevent a
collapse of the WCR because the region over which braking is predicted
to occur is occupied by postshock gas at $T>10^{6\;}$K which we assume
to be mostly ionized and consequently to be subject to a negligible
driving force. However, the strength of radiative braking is dependent
on the \cite{Castor:1975} parameters adopted, i.e. the $k$ and
$\alpha$ \citep{Gayley:1997}. When calculating the line force applied
to the WR's wind by the O star's radiation we use the O star's
\cite{Castor:1975} parameters ($k_{\rm O}=0.3$ and $\alpha_{\rm
  O}=0.52$). Hence, if the WR's radiation-wind coupling was adopted
($k_{\rm WR}=0.42$ and $\alpha_{\rm WR}=0.47$) the decelerative line
force would be greater and radiative braking may be more effective
than in our calculations.

\begin{figure}[!]
  \begin{center}
    \begin{tabular}{cc}
\resizebox{40mm}{!}{\includegraphics{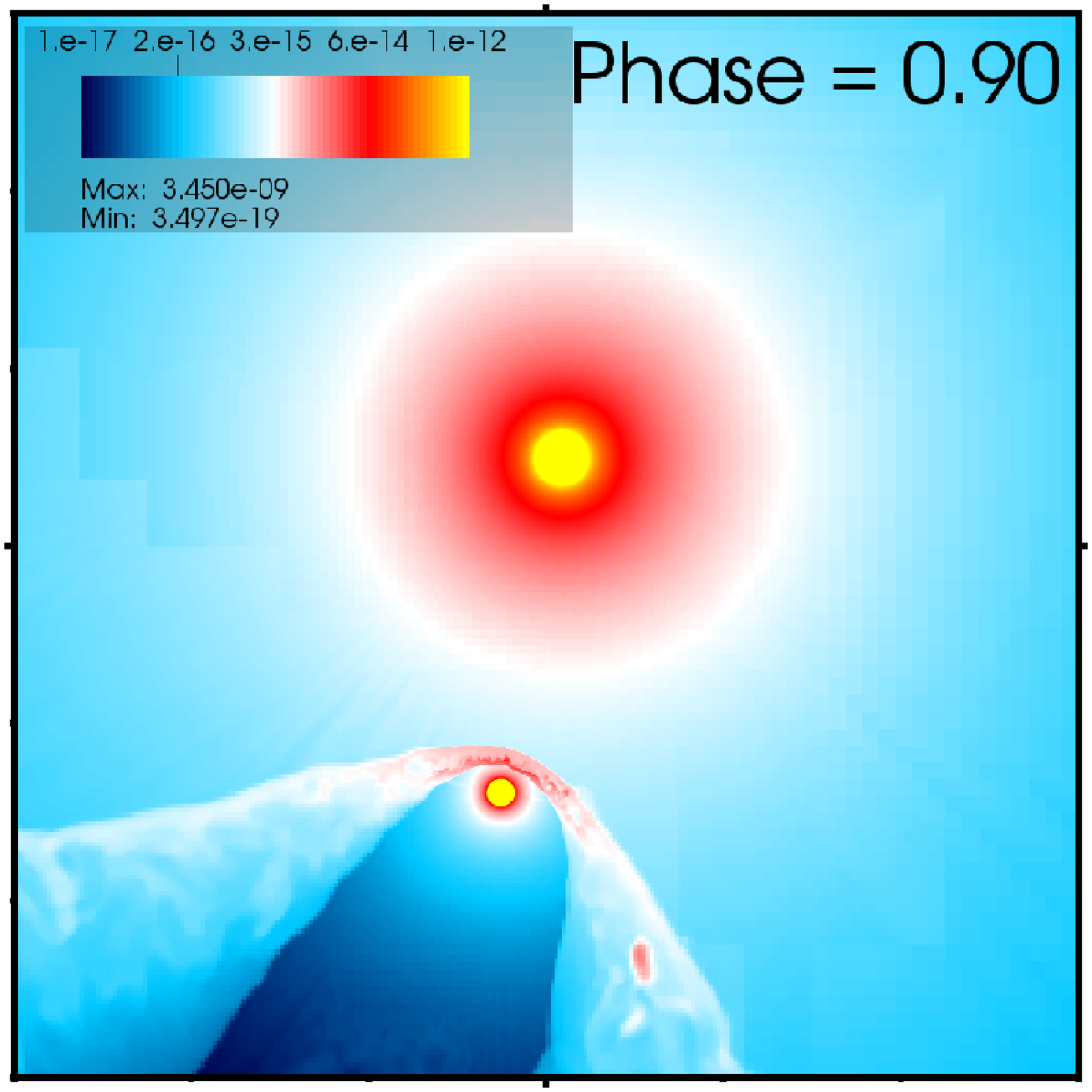}} & 
\resizebox{40mm}{!}{\includegraphics{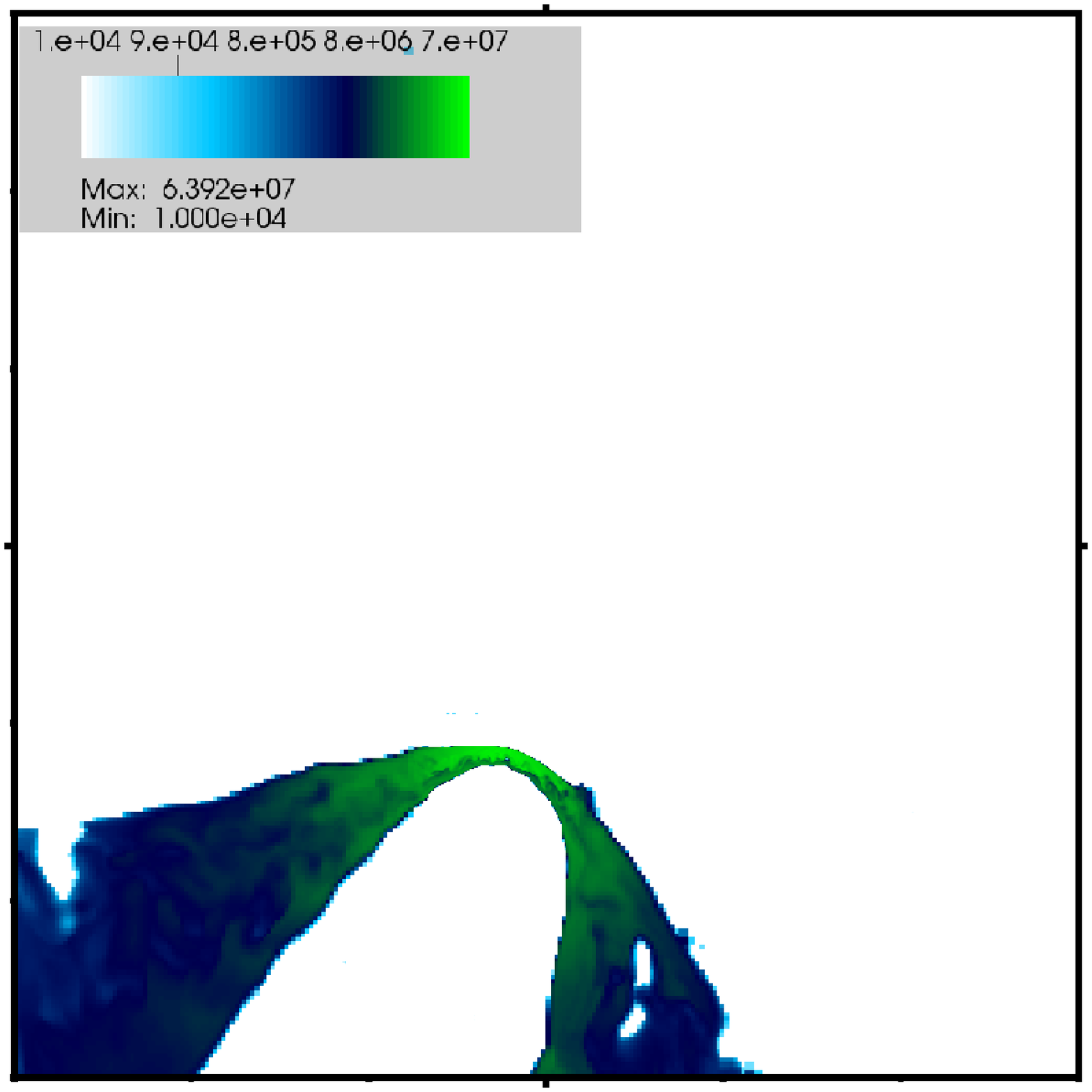}} \\
\resizebox{40mm}{!}{\includegraphics{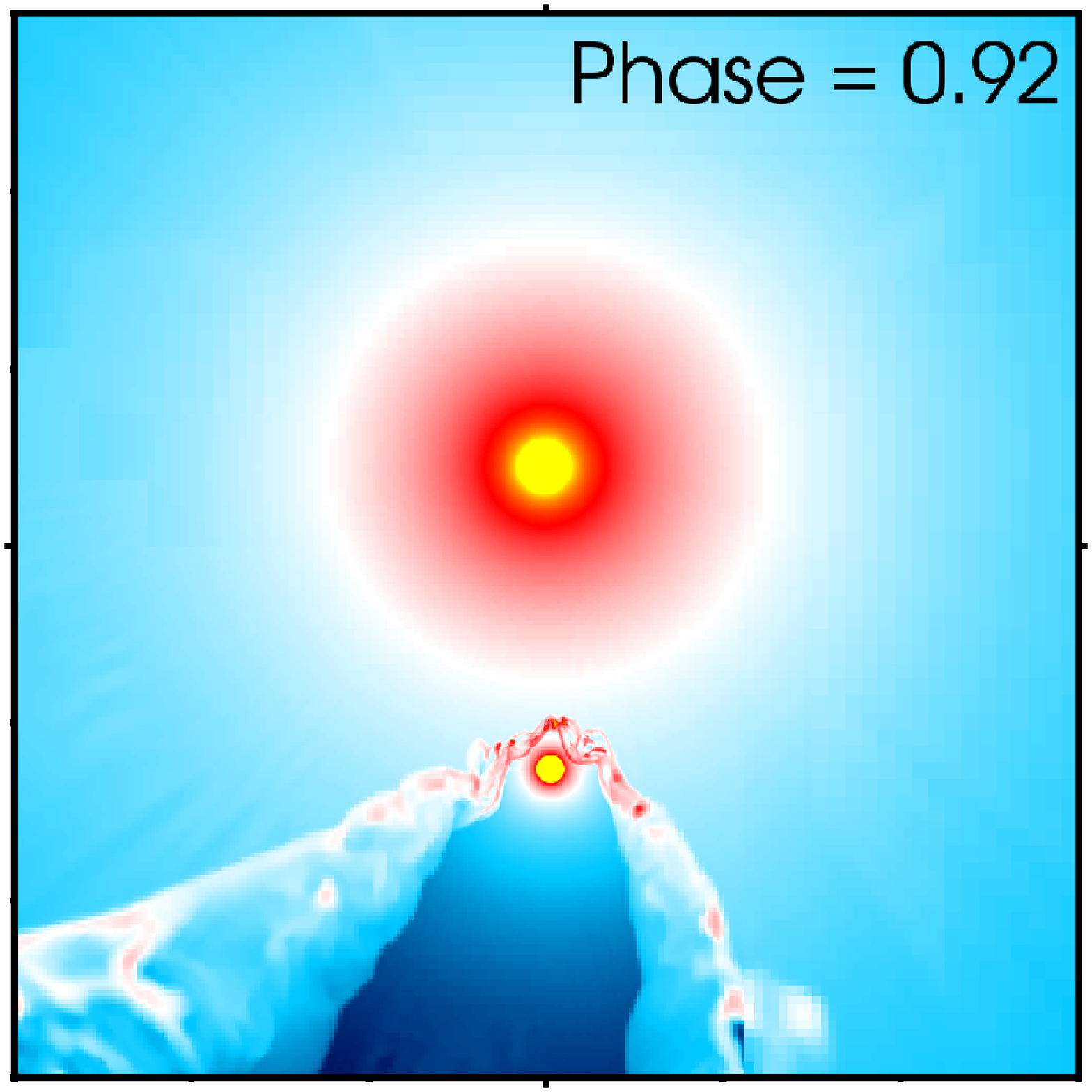}} & 
\resizebox{40mm}{!}{\includegraphics{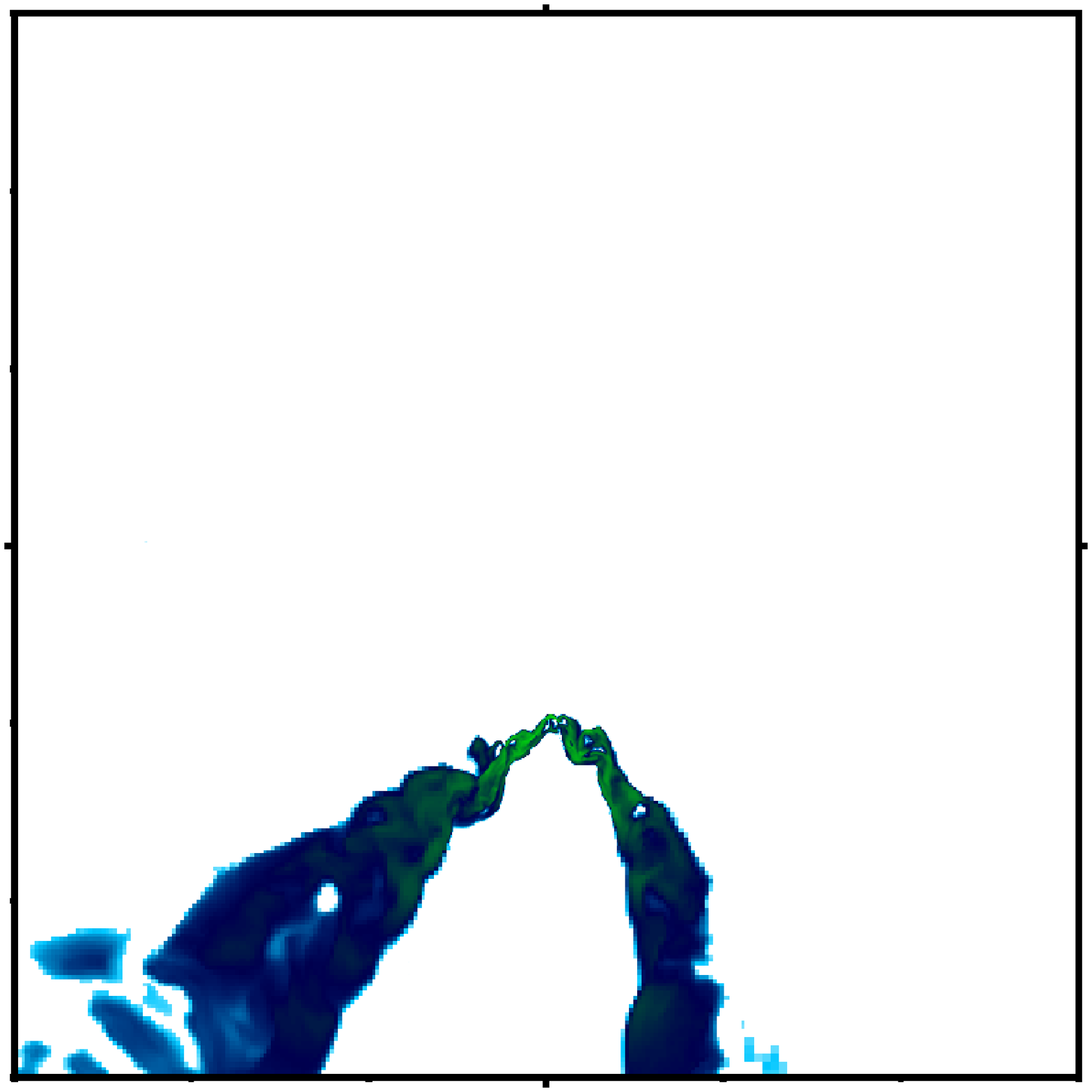}}\\
\resizebox{40mm}{!}{\includegraphics{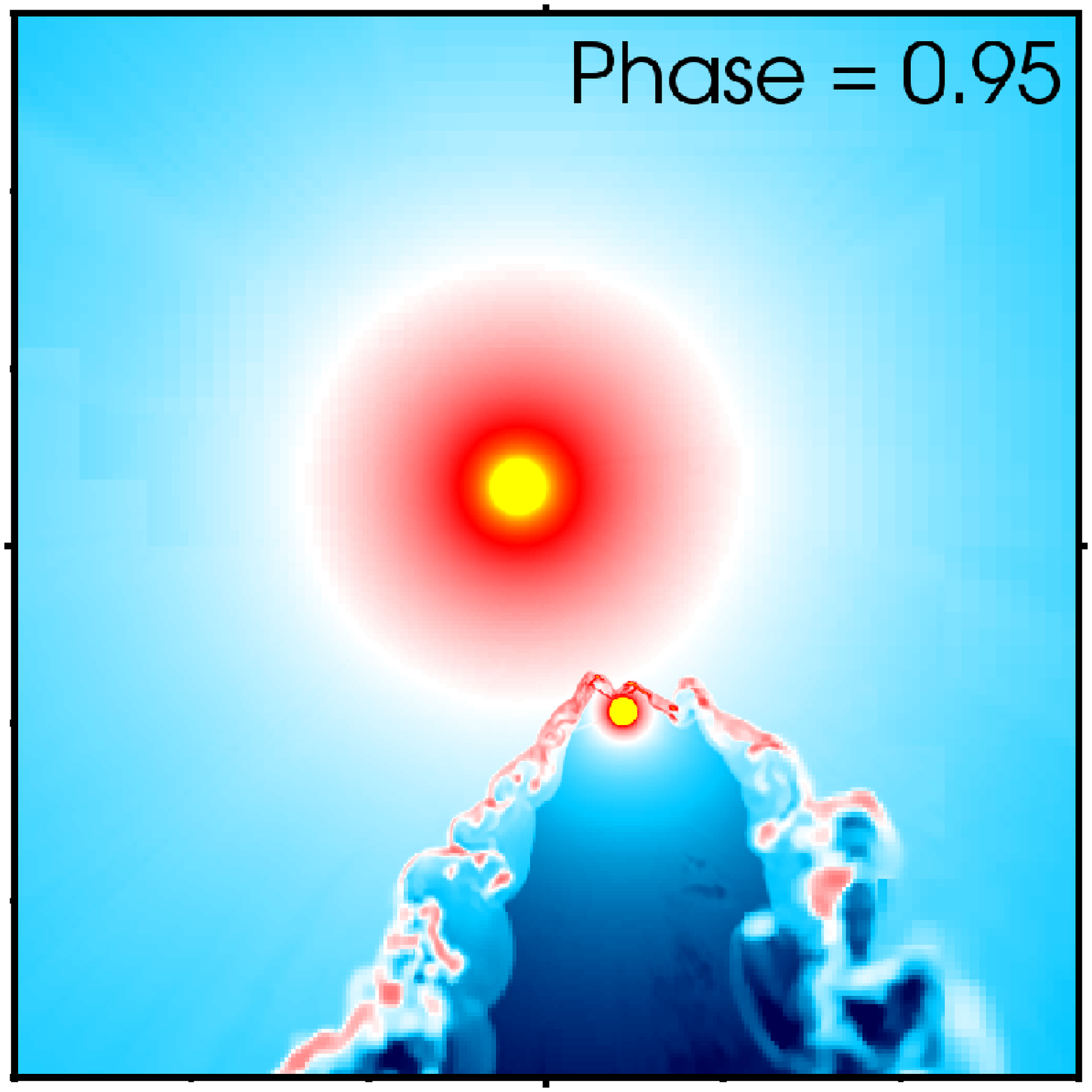}} & 
\resizebox{40mm}{!}{\includegraphics{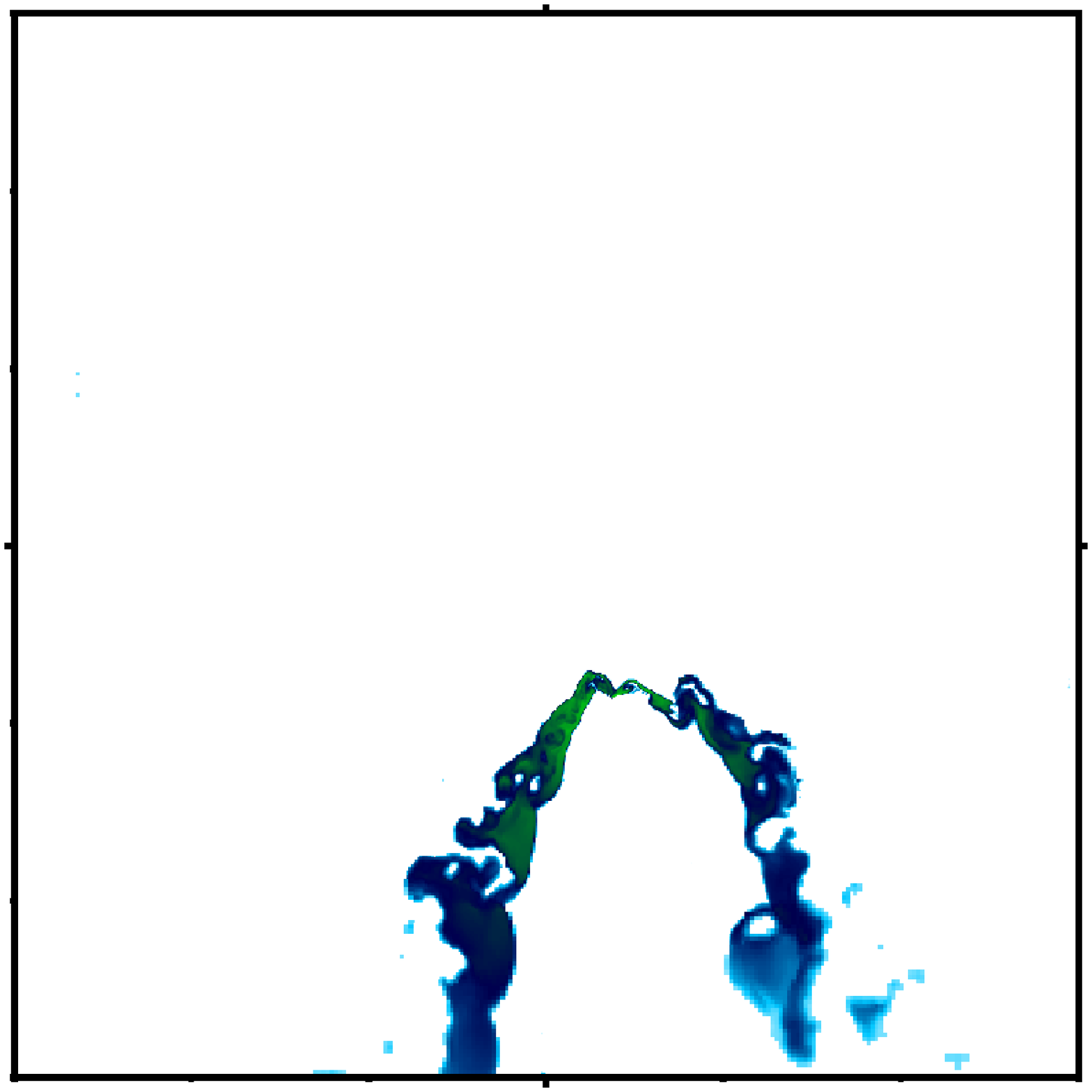}}\\
\resizebox{40mm}{!}{\includegraphics{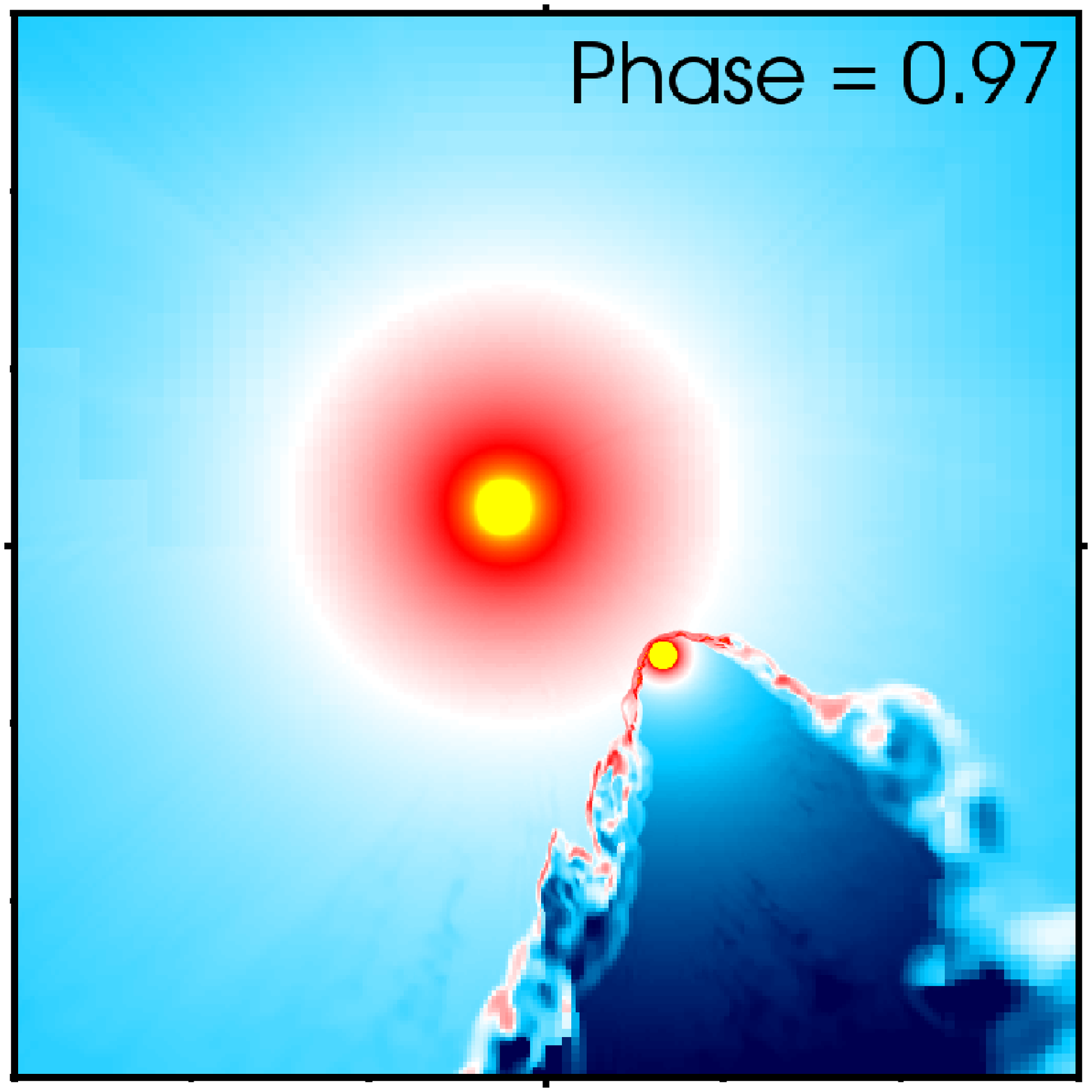}} & 
\resizebox{40mm}{!}{\includegraphics{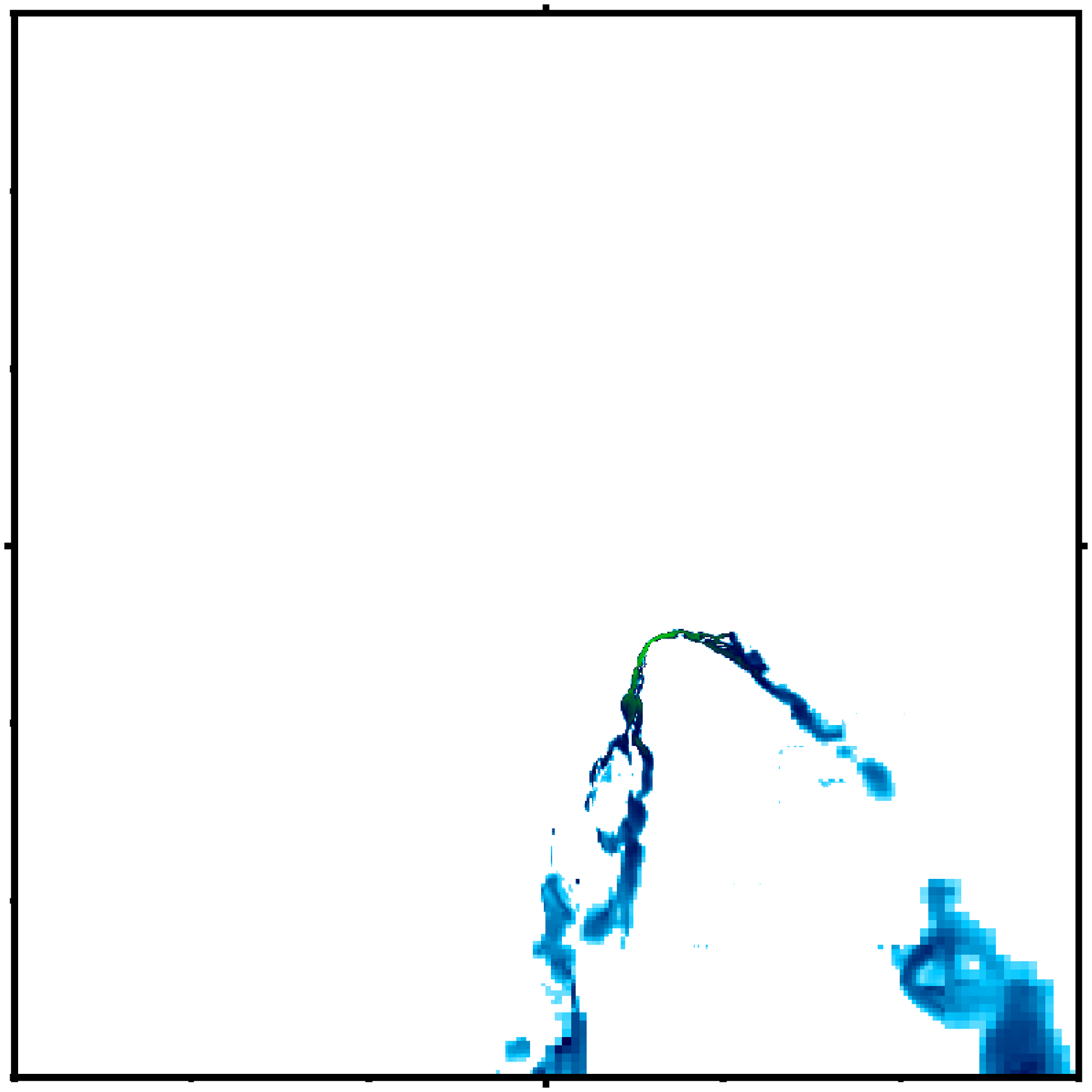}} \\
\resizebox{40mm}{!}{\includegraphics{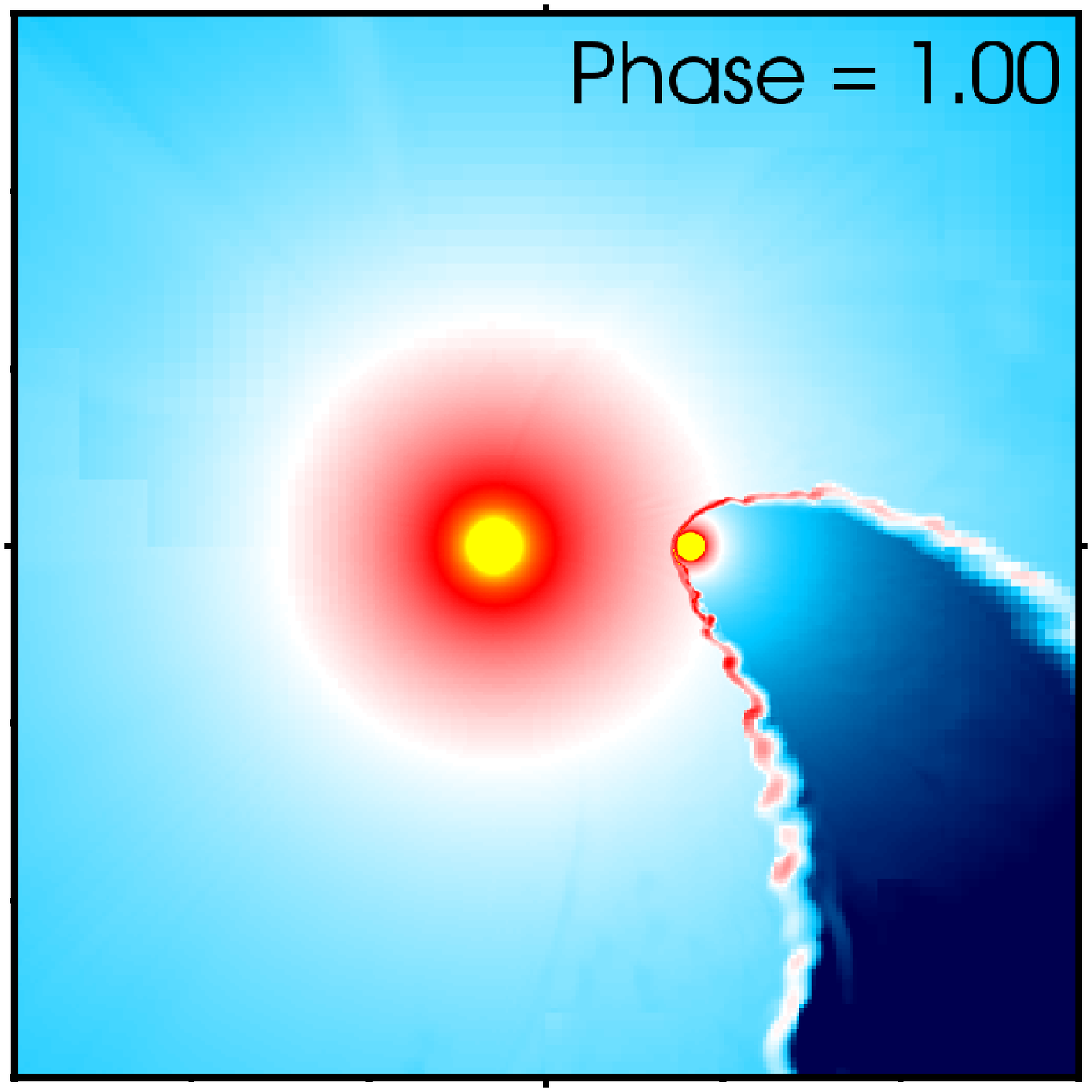}} & 
\resizebox{40mm}{!}{\includegraphics{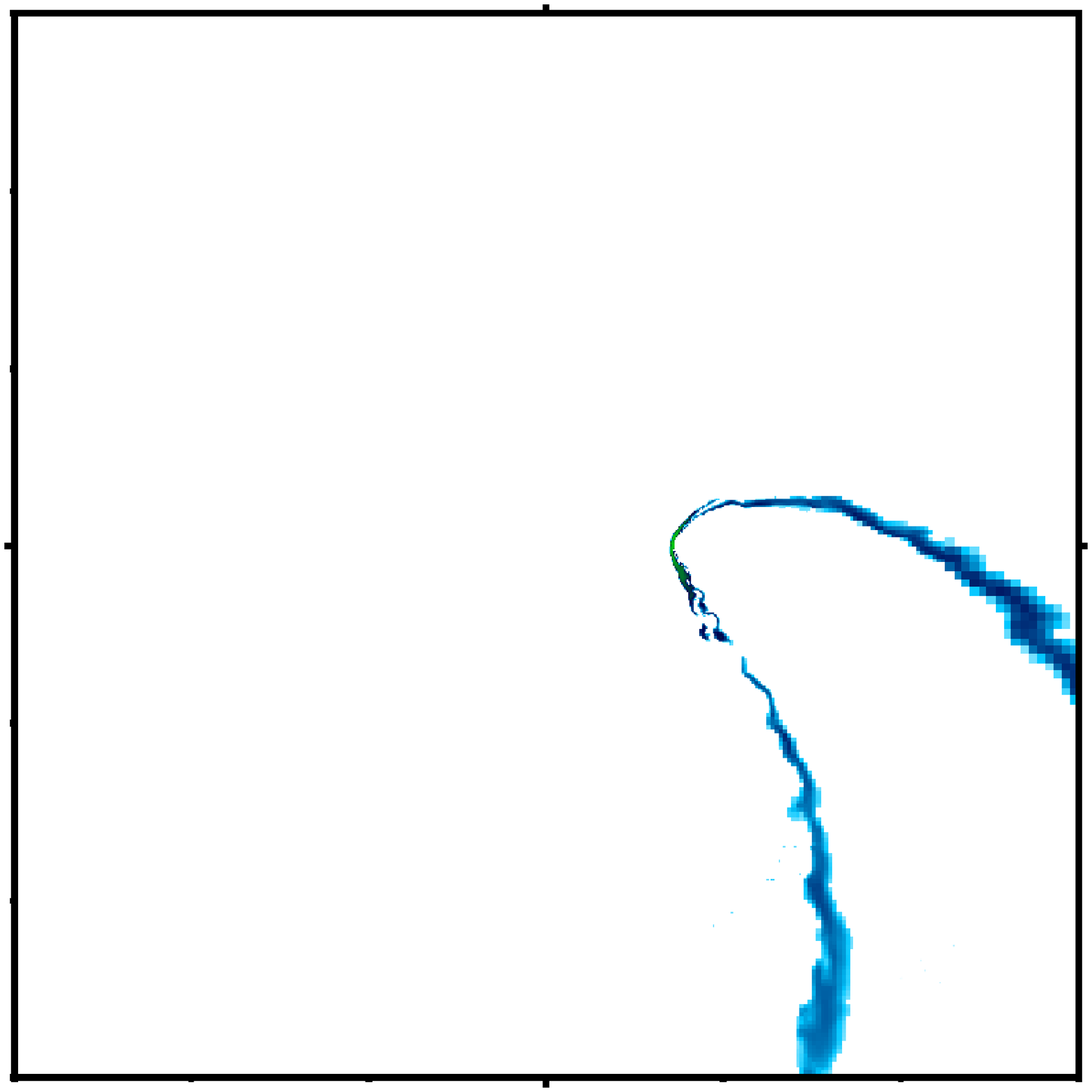}}\\
    \end{tabular}
    \caption{Sequence of snapshots of gas density (left column) and
      temperature (right column) in the orbital ($x-y$) plane
      illustrating the collapse of the WCR onto the O star. From top
      to bottom: $\phi = 0.90$, 0.92, 0.95, 0.97, and 1.00. All plots
      show a region of $\pm3\times10^{13}\;$cm.}
    \label{fig:collapse_images}
  \end{center}
\end{figure}

Interestingly, as the stars reach $\phi=1.0$ the apex of the WCR is
essentially stabilised against NTSIs by the continuing collapse
(i.e. the ram pressure of the WR's wind pins the WCR apex to the O
star preventing oscillations) and only relatively small amplitude KH
instabilities can be seen in the trailing arm of the WCR
(Fig.~\ref{fig:collapse_images}). Despite the O star's wind being
overwhelmed between the stars it continues to drive a wind on the side
facing {\it away from} the WR and, as in model A, carves a tenuous
cavity through the dense WR wind (Fig.~\ref{fig:driven_images}). As
the wind driven from the far side of the O star shocks against the
leading arm of the WCR (which becomes wrapped around the stars) it
reheats the postshock gas, and consequently the leading arm of the WCR
is hotter than the trailing arm (see the temperature plots at
$\phi=1.0\;$and 1.1 in Figs.~\ref{fig:vterm_images} and
\ref{fig:driven_images}).

\begin{figure}[!]
  \begin{center}
    \begin{tabular}{cc}
\resizebox{40mm}{!}{\includegraphics{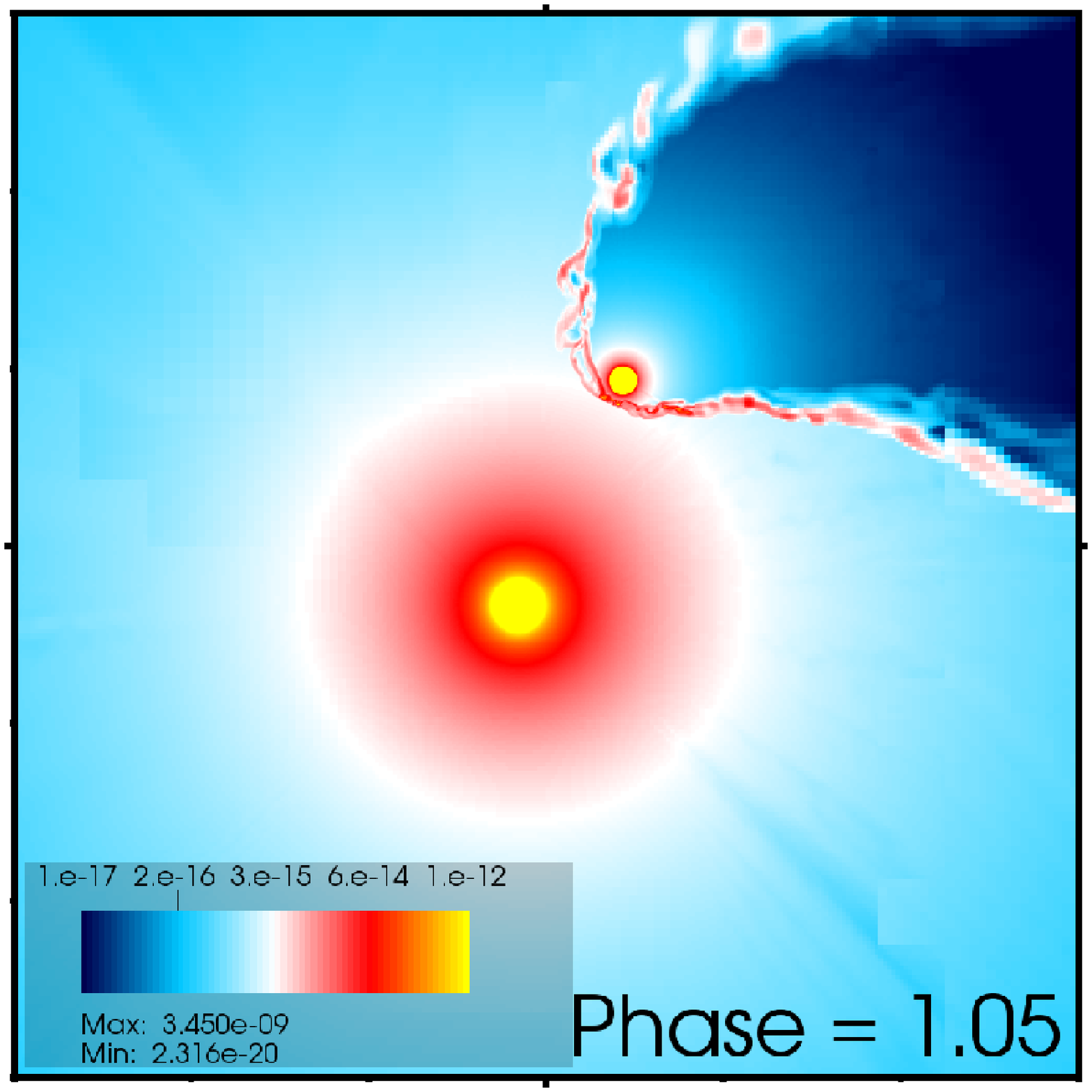}} & 
\resizebox{40mm}{!}{\includegraphics{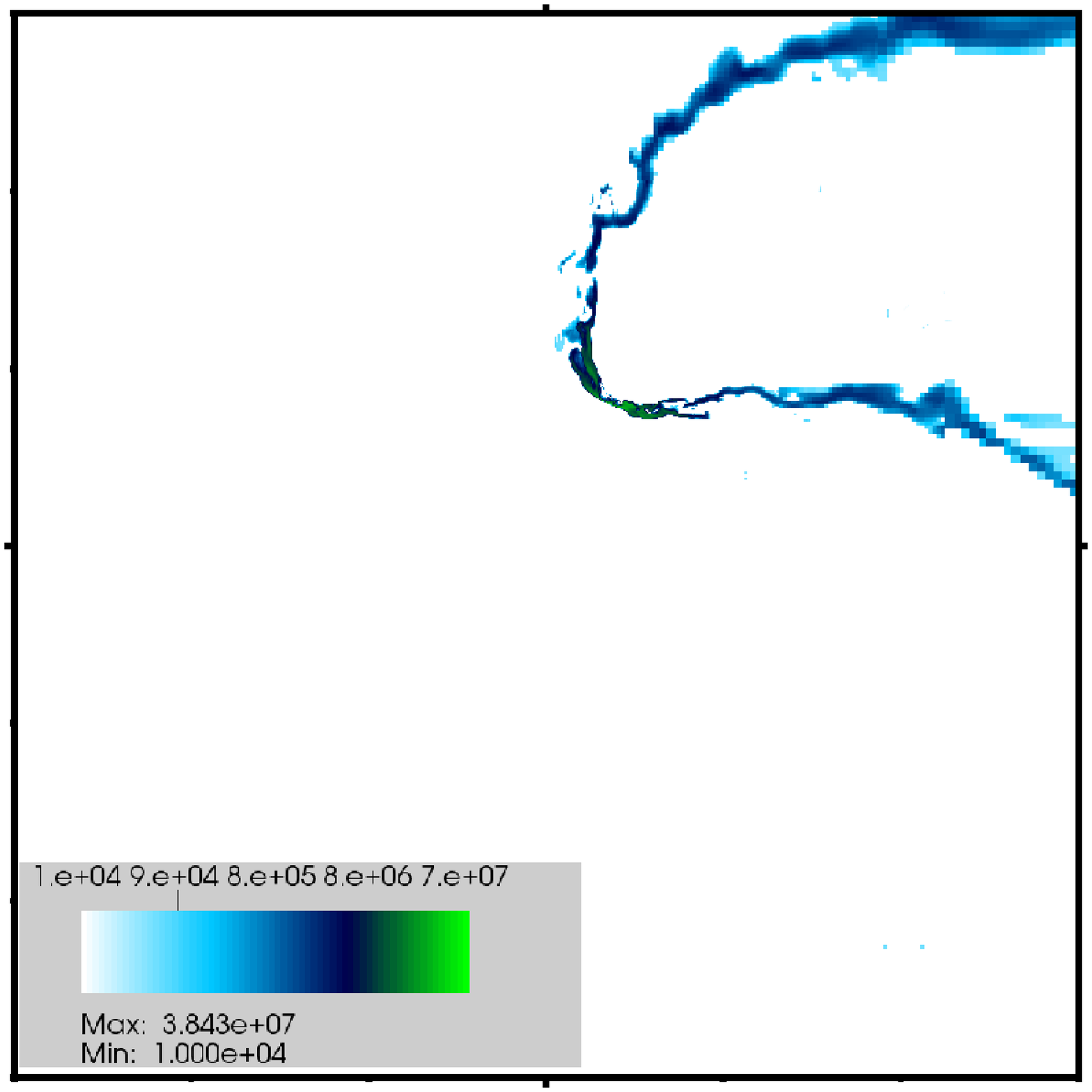}} \\
\resizebox{40mm}{!}{\includegraphics{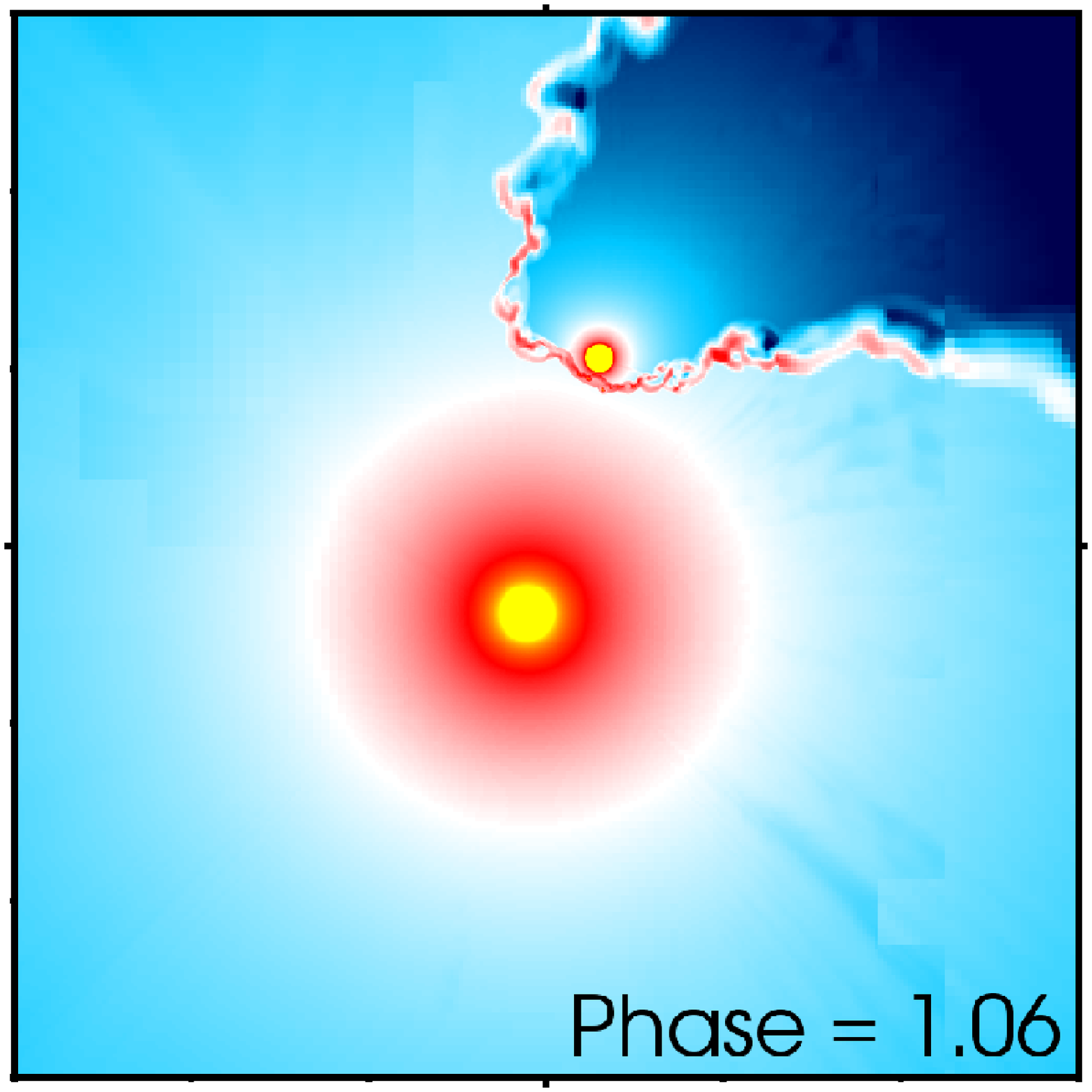}} & 
\resizebox{40mm}{!}{\includegraphics{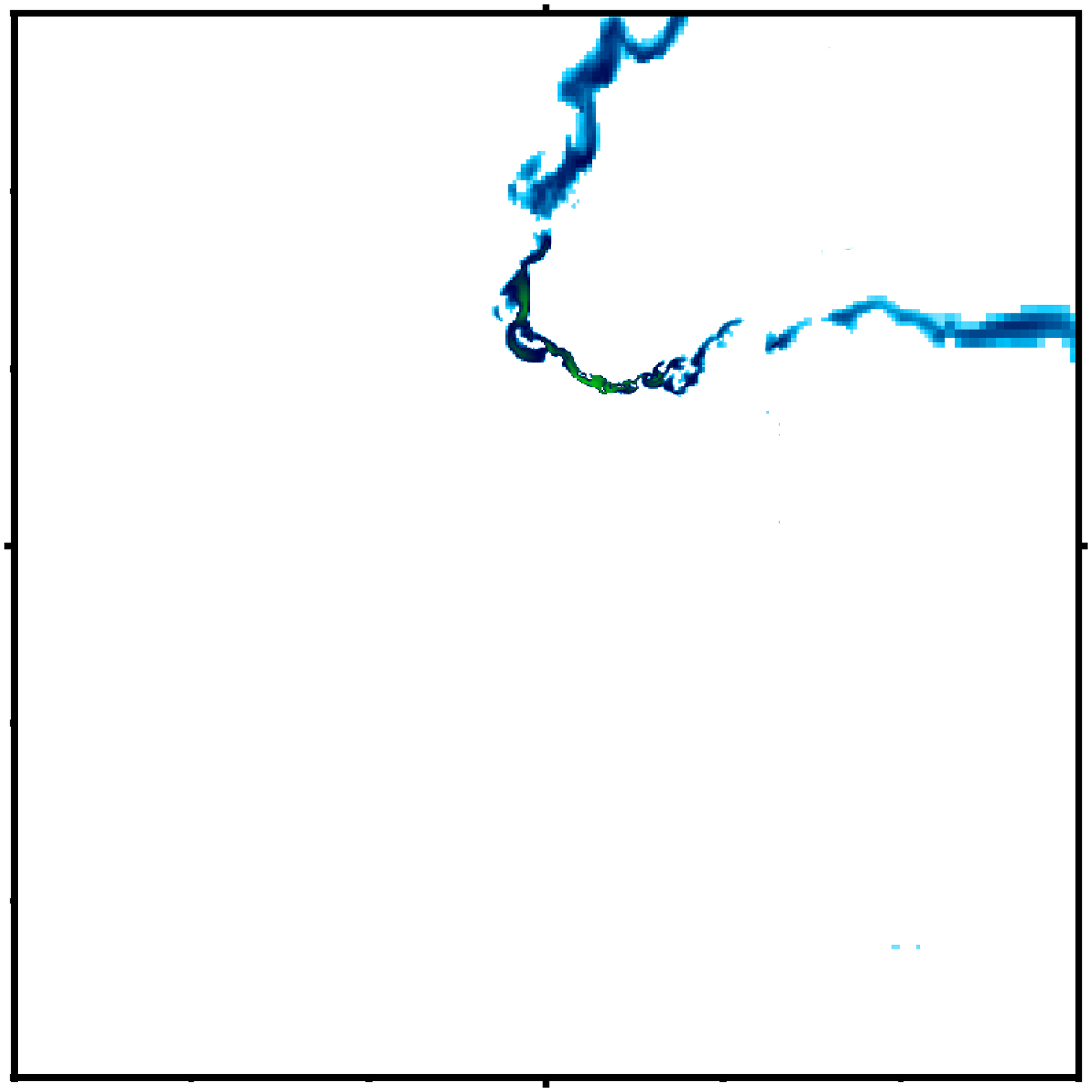}}\\
\resizebox{40mm}{!}{\includegraphics{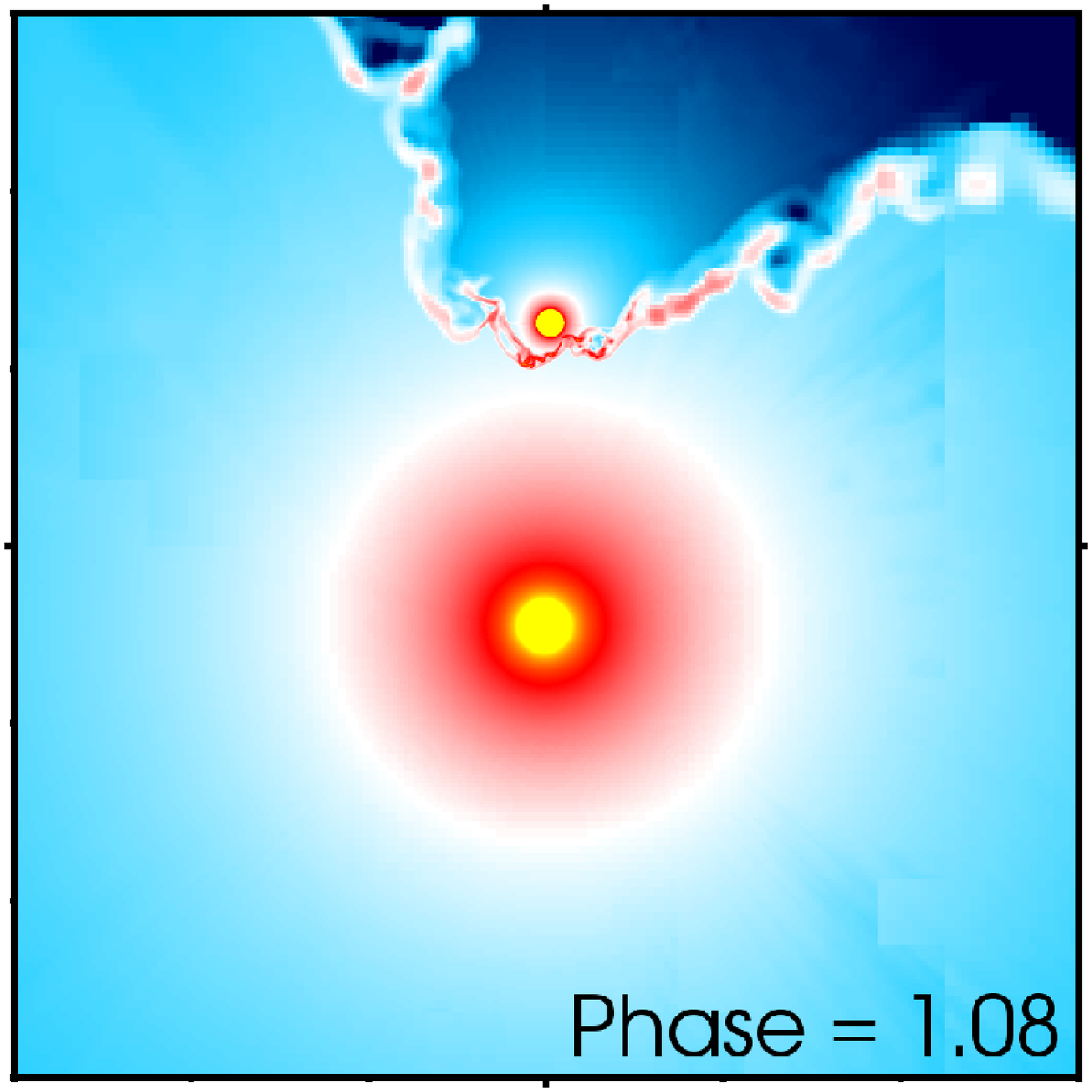}} & 
\resizebox{40mm}{!}{\includegraphics{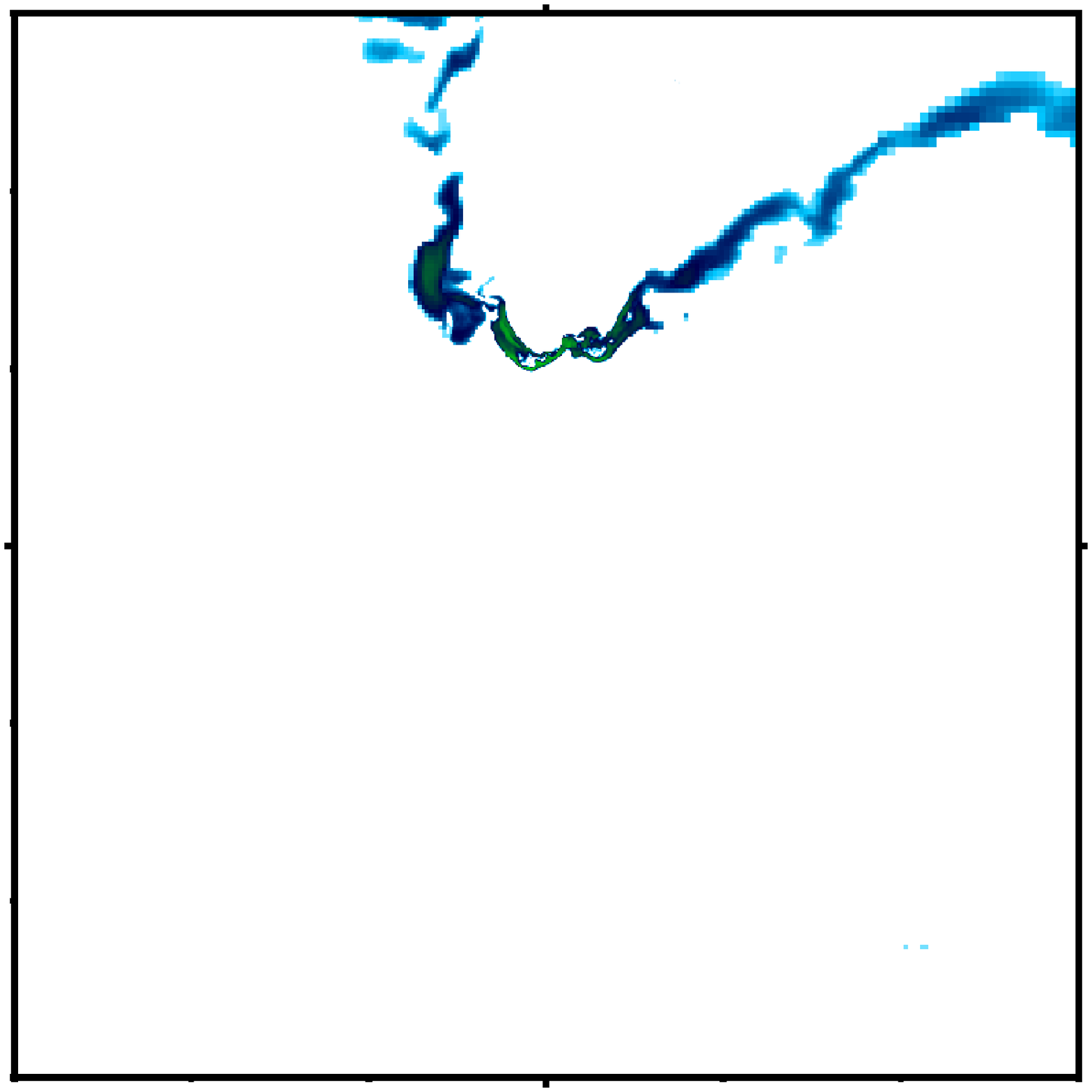}}\\
\resizebox{40mm}{!}{\includegraphics{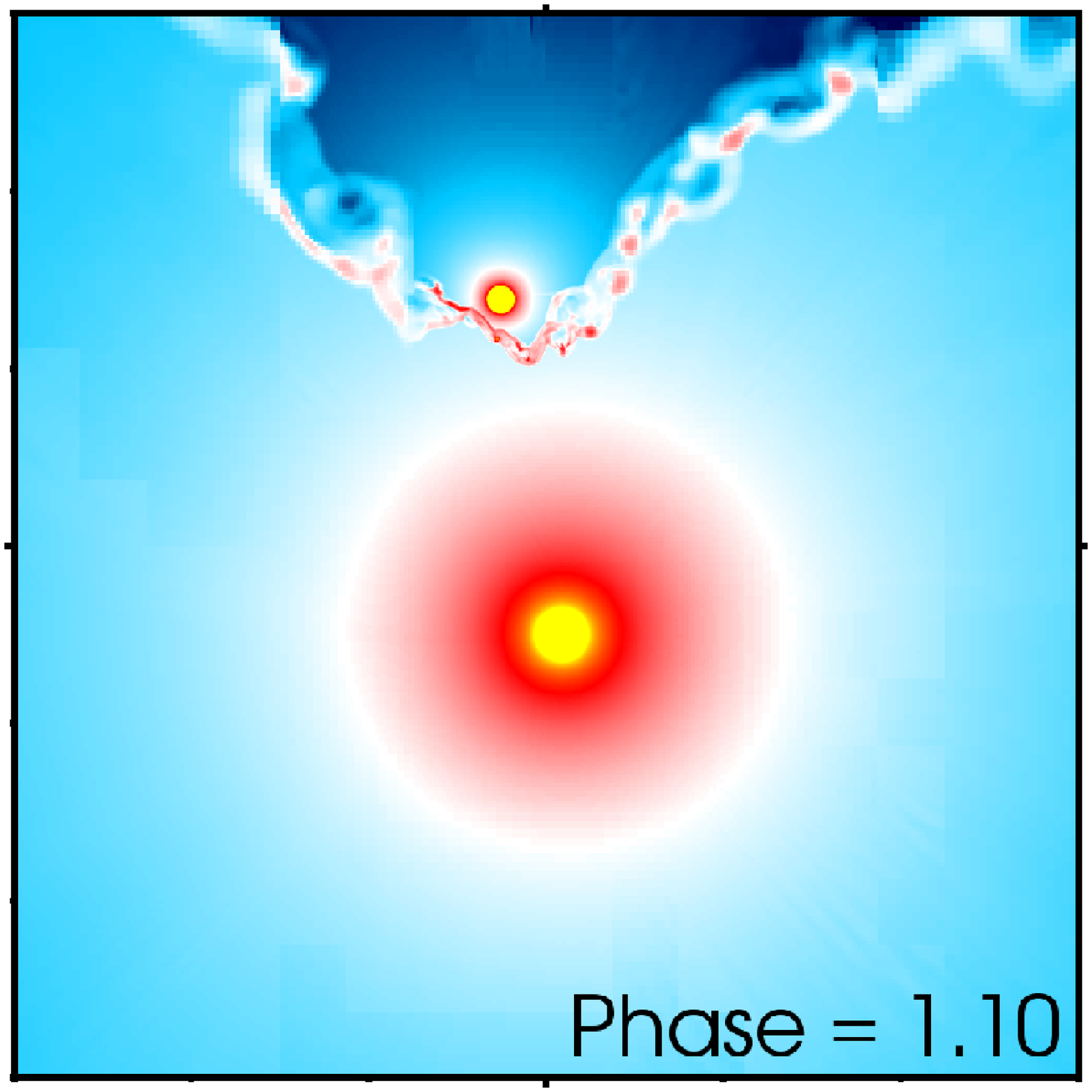}} & 
\resizebox{40mm}{!}{\includegraphics{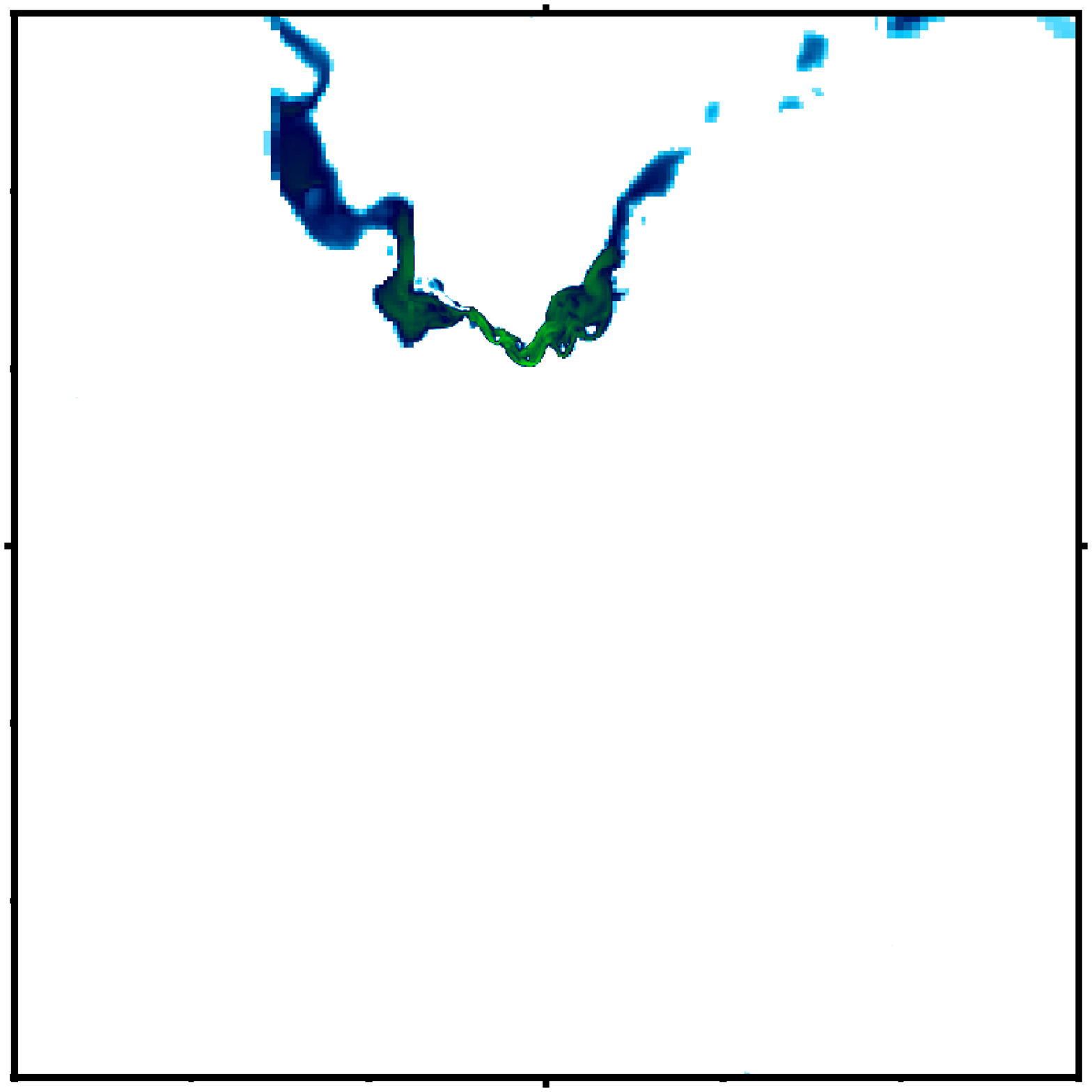}} \\
    \end{tabular}
    \caption{Same as Fig~.\ref{fig:collapse_images} except showing the
      recovery of the WCR from collapse. From top to bottom: $\phi =
      1.05$, 1.06, 1.08, and 1.10. All plots show a region of
      $\pm3\times10^{13}\;$cm.}
    \label{fig:recovery_images}
  \end{center}
\end{figure}

Following periastron, as the separation of the stars expands, the
preshock ram pressure of the WR's wind - which pins the shocks onto
the O star - decreases. Therefore, when it weakens sufficiently, the
WCR is permitted to move away from the O star. Simultaneously, as the
stars recede, radiative inhibition decreases and the acceleration of
the O star's wind increases. The O star shock is buried so deeply in
its wind acceleration region that a small difference in its position
considerably alters the preshock velocity. Hence, as soon as the O
star's wind is given some room to accelerate the acquired preshock ram
pressure greatly increases, helping the WCR to separate from the O
star. Subsequently, the higher preshock velocities lead to higher
postshock gas temperatures/pressures, which increase the stability of
the WCR apex against NTSIs. The snapshots in
Fig.~\ref{fig:recovery_images} show the gradual transition from
collapse to recovery. Notice that the recovery is not instantaneous
and, as with the collapse, NTSIs vigorously disrupt the WCR and
sporadic collisions of the WCR against the O star occur
(e.g. $\phi=1.08\;$in Fig.~\ref{fig:recovery_images}). However, as the
amplitude of the oscillations increases, the O star's wind is allowed
to accelerate to higher velocities before it shocks and, as such, the
postshock gas pressure sees a commensurate increase. Following $\phi
=1.10$ the WCR maintains a clear separation from the O star.

On the basis of the wind-wind momentum balance alone, the separation
of the WCR from the O star is inevitable (Fig.\ref{fig:chi}). However,
by initiating the winds in model B within a radial distance of $\sim
1.15\;{\rm R_{\ast}}$ \citep[where $0.15\;{\rm R_{\ast}} \ge 3\;$cells
  -][]{Pittard:2009} we are essentially enforcing a lower limit to the
wind ram pressure (i.e. the O star's wind velocity never reaches zero)
and influencing the time at which the shocks detach from the O
star. For instance, in models of Iota Orionis, \cite{Pittard:1998b}
found that the orbital phase at which the WCR recovered was tied to
the size of the annulus used to initiate the stellar wind with a
smaller annulus leading to a more prolonged collapse. In reality, the
incoming wind may interfere with the initiation of the wind and
prolong a collapse/disruption of the WCR. Hence, although the exact
time at which the WCR separates from the O star may be resolution
dependent, its occurrence in model B is qualitatively correct.

We note that numerical heat conduction between hot tenuous gas and
cool dense gas will occur in the simulations
\citep{Parkin_Pittard:2010}. In model B this will affect the
temperature of postshock gas, and hence the rate at which it cools. For
instance, during the collapse of the WCR the postshock WR wind will be
caused to cool more rapidly and the O star wind will be artificially
heated, potentially to X-ray emitting temperatures. This effect is
also present in model A, but to a lesser extent due to the lower
contrasts between adjacent postshock gas (i.e. there are cool clumps
in the postshock gas in model B but not in model A). However, the
conclusions drawn from this work are not affected by the undesirable
influence of numerical heat conduction.

\subsection{X-ray emission}

The collision of the winds in WR22 generates hot ($T>10^{7}\;$K)
plasma which emits radiation at X-ray wavelengths, providing a direct
probe of the postshock winds, and an indirect probe of the preshock
winds. However, a recent analysis of {\it XMM-Newton} observations of
WR22 by \cite{Gosset:2009} revealed a number of discrepancies between
predictions from a wind-wind collision model and observations. To
determine whether the contrasting flow dynamics from models A and B
can shed any light on these issues we have performed X-ray radiative
transfer calculations on the simulation output, details of which can
be found in \S~\ref{subsec:xray_emission}. We define our line of sight
geometry as follows: the inclination angle, $i$, is measured against
the $z$-axis ($i=0^{\circ}$ would view the system from directly above
the orbital plane), and the angle $\theta$ is measured against the
{\it negative} $x$-axis (O star in front at apastron) such that
$\theta$ increases in the prograde direction ($\theta =90^{\circ}$
would align the line of sight with the negative $y$-axis). We adopt viewing
angles of $i=85^{\circ}$ and $\theta = 0^{\circ}$, in agreement with
the system orientations determined by \cite{Rauw:1996} and
\cite{Schweickhardt:1999}.

\subsubsection{Lightcurves}

\begin{figure}
  \begin{center}
    \begin{tabular}{c}
\resizebox{80mm}{!}{\includegraphics{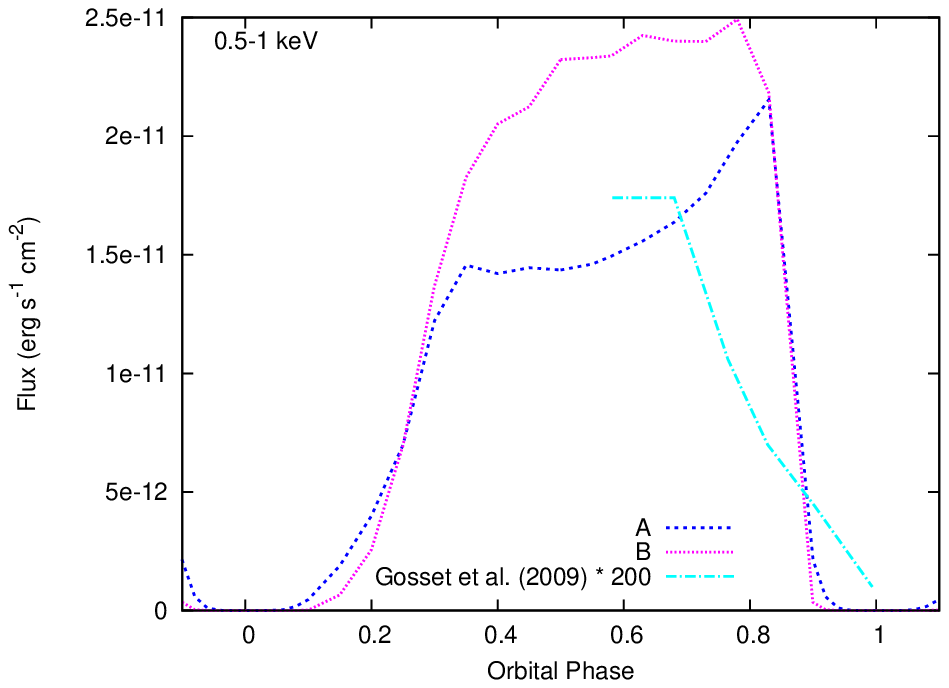}}  \\
\resizebox{80mm}{!}{\includegraphics{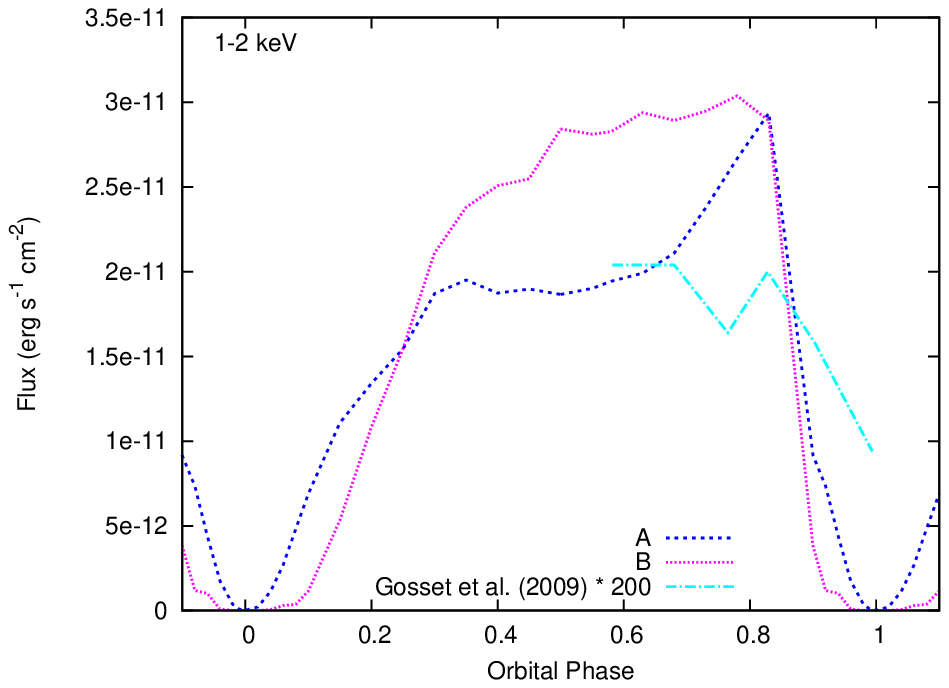}}  \\
\resizebox{80mm}{!}{\includegraphics{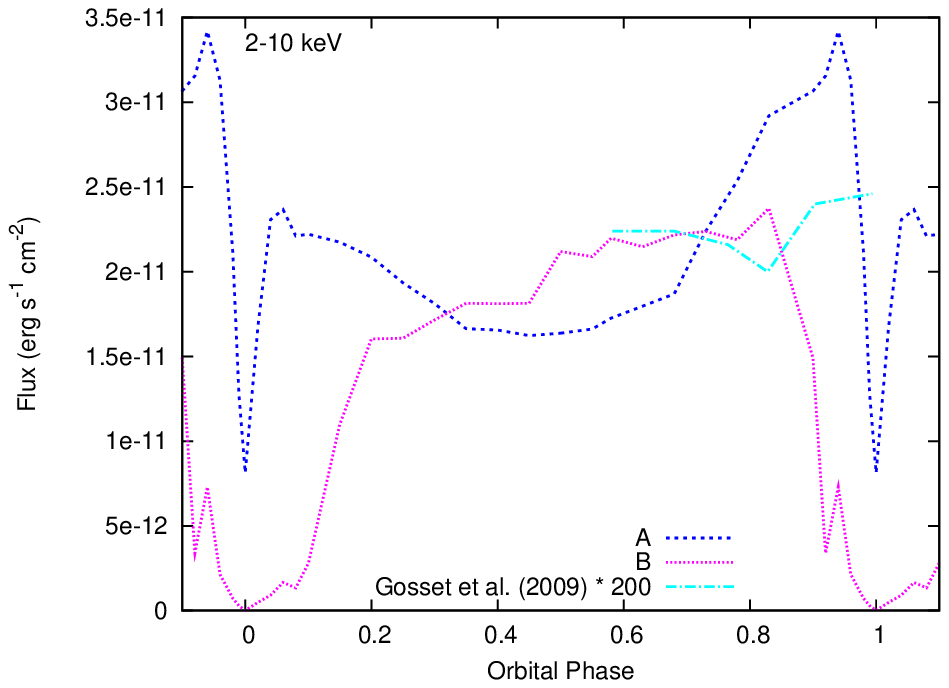}}  \\
    \end{tabular}
    \caption{X-ray lightcurves calculated from models A and B in the
      0.5-1 (upper panel), 1-2 (middle panel), and 2-10~keV (lower
      panel) energy bands. The values derived by \cite{Gosset:2009}
      from spectral fits to {\it XMM-Newton} spectra are plotted for
      comparison - note that they have been rescaled (upwards) by a
      factor of 200 for illustrative purposes. Note the difference in
      scale between the plots.}
    \label{fig:lc}
  \end{center}
\end{figure}

The 0.5-1, 1-2, and 2-10~keV X-ray lightcurves calculated from models
A and B are shown in Fig.~\ref{fig:lc}. Although the separate
contributions are not shown, the X-ray emission is dominated by the
postshock WR wind due to its higher density and mean atomic
weight. Examining the lightcurves for model A one sees a reasonably
constant flux between $\phi \simeq 0.35-0.65$ corresponding to the
relatively slow rate of change of the binary separation whilst the
stars are close to apastron. As the stellar separation contracts, the
preshock, and thus postshock, gas density increases resulting in a
rise in the observed flux. However, approaching periastron, lines of
sight to the X-ray emitting plasma begin to pass through the WR's wind
and the abrupt increase in absorption causes the observed flux to
decline. The turn-over occurs at an earlier phase for X-rays with
$E<2\;$keV due to the higher susceptibility to absorption compared to
those at $E>2\;$keV. The decline in observed flux continues until a
minimum is reached at periastron passage thence attenuation reaches a
maximum. The spatial extent of the X-ray emission region means that an
eclipse by the WR star will have only a small impact, hence the dip in
observed flux around periastron mainly occurs due to absorption by the
unshocked WR wind \citep{Parkin:2008}.

The model B lightcurves differ from those of model A in a number of
ways. Firstly, the observed flux at apastron in model B is higher than
in model A due to the distance between the WR and its respective shock
being slightly smaller in the former (see
\S~\ref{subsec:modelB}). Secondly, the decline in flux begins at the
same orbital phase ($\phi\simeq0.8$) in the 0.5-1, 1-2, and 2-10~keV
lightcurves. Interestingly, in model B the start of the decline in the
2-10 keV flux corresponds to the initiation of the WCR disruption,
unlike in model A where no disruption occurs. Similarly, the egress in
the observed flux following periastron is tied to the recovery of the
WCR. The highly unstable WCR introduces flare-like features into the
lightcurves. For example, during the rapid decline in the observed
flux ($\phi\simeq 0.8-1.0$) there is a sudden spike at $\phi=0.94$
caused by the WCR oscillating away from the O star
(Fig.~\ref{fig:collapse_images}). It is interesting to note that
flare-like features are observed in the X-ray lightcurve of
$\eta~$Carinae \citep{Corcoran:2001, Corcoran:2005,
  Corcoran:2010}. Although \cite{Moffat:2009} dismissed instabilities
in the WCR as the mechanism responsible for the flare-like features,
the model B lightcurve provides some evidence to the contrary.

The most apparent difference between models A and B is the drastic
reduction in the 2-10~keV flux caused by the disruption, and
subsequent collapse onto the O star, of the WCR in the
latter. Consequently, at periastron the difference between the two
models reaches a maximum with the 2-10~keV flux from model B being a
factor of $\simeq18$ lower than from model A. This occurs because
during the collapse the fraction of the WR's wind being shocked is
approximately the fractional solid angle subtended by the O star,
which is much lower than occurs when a stable WCR exists.

For comparison, we also plot the X-ray fluxes derived from {\it
  XMM-Newton} observations of WR22 by \cite{Gosset:2009} in
Fig.~\ref{fig:lc}. Clearly, the models over-predict the
\citeauthor{Gosset:2009} values in all three energy bands (note that
the \citeauthor{Gosset:2009}~\citeyear{Gosset:2009} values have been
rescaled in Fig.~\ref{fig:lc}). At apastron, when a stable WCR exists
in both models, the 0.5-1, 1-2, and 2-10~keV fluxes are over-estimated
by factors of $\simeq 220$, 230, and 170, respectively. The major
concern is the over-estimate in the 2-10~keV flux, as X-rays in this
energy band are less susceptible to absorption and, therefore, the
discordance between the models and observations cannot be attributed
to insufficient attenuation. Hence, to reduce the observed 2-10~keV
flux requires a reduction in the {\it intrinsic} flux, i.e. an
alteration to the WCR. Encouragingly, the disruption/collapse of the
WCR closes the gap considerably; at $\phi=0.994$ model B only
over-estimates the \cite{Gosset:2009} values by factors of $\simeq
7.8$, 4.7, and 6.8 in the 0.5-1, 1-2, and 2-10~keV energy bands,
respectively. The comparison of the model lightcurves and the
\cite{Gosset:2009} values would therefore suggest that the WCR must be
significantly disrupted/collapsed onto the O star throughout the
entire orbit. We consider this point further, along with revisions to
parameters to facilitate it, in \S~\ref{subsec:revised_model}.

\subsubsection{Column density}
\label{subsec:column_density}

\begin{figure}
  \begin{center}
    \begin{tabular}{c}
\resizebox{80mm}{!}{\includegraphics{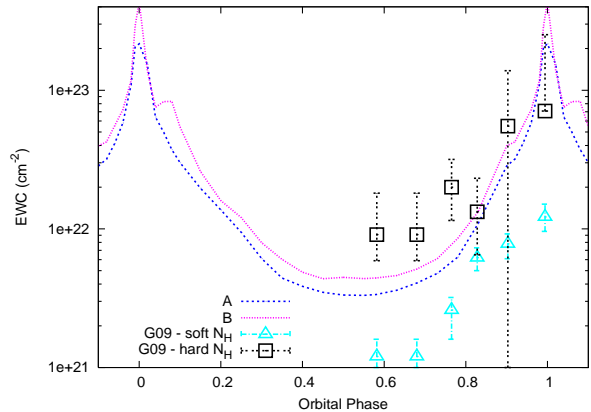}}  \\
    \end{tabular}
    \caption{Emission weighted column density (EWC) calculated from
      models A and B. The values derived by \cite{Gosset:2009} from
      spectral fits to {\it XMM-Newton} spectra are also plotted (see
      \S~\ref{subsec:column_density}), with the formal error bars
      attained from the spectral fitting. The plotted values do not
      include the column density due to the interstellar medium
      ($2.5\times10^{21}\;{\rm cm^{-2}}$).}
    \label{fig:EWC}
  \end{center}
\end{figure}

\begin{figure}
  \begin{center}
    \begin{tabular}{c}
\resizebox{80mm}{!}{\includegraphics{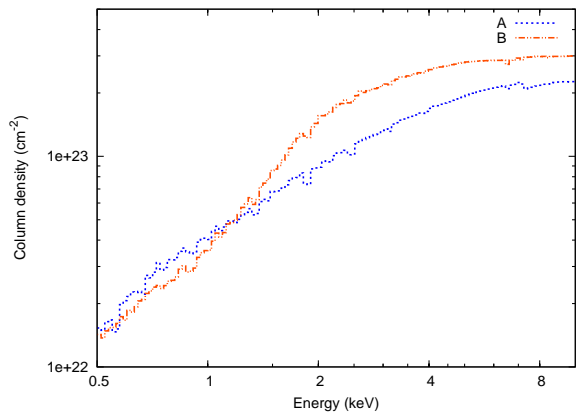}}  \\
    \end{tabular}
    \caption{Emission weighted column density (EWC) as a function of
      energy calculated from models A and B at $\phi=0.994$. The
      column density due to the interstellar medium
      ($2.5\times10^{21}\;{\rm cm^{-2}}$) is additional to the
      plotted values.}
    \label{fig:nh_spec}
  \end{center}
\end{figure}

The emission-weighted column densities (EWCs)\footnote{The EWC is
  calculated as $N_{\rm EWC}= \Sigma N_{\rm H} L_{\rm int X} / \Sigma
  L_{\rm int X} $, where $N_{\rm H}$ and $L_{\rm int X}$ are the
  column density and 0.5-10~keV intrinsic luminosity from a given line
  of sight, and the summation is over all sight lines (pixels) in the
  X-ray image \citep[e.g.][]{Parkin:2008, Parkin:2009b}.} calculated
from models A and B are plotted in Fig.~\ref{fig:EWC}. A minimum in
the column density occurs for both models at $\phi=0.5$ corresponding
to the stars at apastron and the lines of sight to the X-ray emitting
plasma passing through the O star wind. The column density then rises
gradually between $\phi=0.5-0.8$ as the WCR, and thus lines of sight
to the X-ray emitting plasma, move deeper into the O star wind. At
$\phi=0.8$ the column density begins to rise rapidly due to the WCR
rotating and some lines of sight now viewing the X-rays through the WR
wind. The column density peaks at periastron when the WCR is at its
deepest point of immersion in the dense WR wind, following which there
is a decrease as the stars recede and lines of sight to the X-ray
emitting plasma traverse progressively lower density wind
material. Comparing the present model column densities to those
computed with the model of \cite{Gosset:2009}, a good agreement is
observed.

The general morphology of the column densities calculated from models
A and B are largely similar for most of the orbit, however, there are
some noticeable differences. The column density is higher for model B
compared to model A at $\phi=0.5$ due to the wind density being
slightly higher for the former in the unshocked O star wind
immediately adjacent to the WCR apex (where the bulk of the X-ray
emission originates from) - this is a numerical artifact of the
different ways by which the winds are driven in the two
simulations. The step in column density in model B which extends to
$\phi\simeq 1.08$ corresponds to lines of sight to the X-ray emitting
plasma passing through cool, dense postshock gas (see
Fig.~\ref{fig:recovery_images}). Interestingly, a similar feature is
observed in simulations of $\eta\;$Carinae \citep{Parkin:2009,
  Parkin:2011b}. As the O star wind recovers, the postshock gas
temperature suddenly increases and cooling becomes less important,
thereby reducing the postshock gas density. Simultaneously, the
rotation of the arms of the WCR results in fewer lines of sight to the
X-ray emitting plasma being obscured by dense postshock gas.

For comparison, two sets of the column densities derived by
\cite{Gosset:2009} are also plotted in Fig.~\ref{fig:EWC}: the lower
values correspond to fits performed with a single column density to
account for all absorption \citep[G09 - soft $N_{\rm H}$ - see table 7
  of][]{Gosset:2009}, whereas the higher values correspond to the
harder component in a two-temperature fit with separate column
densities \citep[G09 - hard $N_{\rm H}$ - see table 6
  of][]{Gosset:2009}. We note that the ``soft'' column densities are
representative of values derived by \cite{Gosset:2009} to the lower
temperature plasma component in two-temperature, two-column density
fits. The observed column density to the hotter emission component is
higher than to the softer plasma component \citep[although these
  values could be lower - see][]{Gosset:2009}, the physical
interpretation of which would be that the hotter plasma resides closer
to the WCR apex and hence lines of sight pass through denser wind
material. In contrast, the lower column density accrued along lines
of sight to the cooler temperature plasma suggests that this emission
comes from a more extended region which is largely viewed through less
dense wind material. Unsurprisingly, the column densities calculated
from models A and B lie roughly between the two sets of observed
column densities as they sample a range of values encountered when
viewing a spatially extended, energy dependent emission
region. Fig.~\ref{fig:nh_spec} shows the column density as a function
of energy at $\phi=0.994$ - the column density at 10 keV is an order
of magnitude higher than at 0.5 keV. This highlights a difficulty when
applying simple spectral fitting models to complicated data, namely
that observationally derived values may be somewhat misleading
\citep{Antokhin:2004, Pittard_Parkin:2010}. For example, based on an
over-prediction of the observed column density, \cite{Gosset:2009}
stated that a more extended emission region was required to reduce the
model column density and attain a better agreement. Yet, as
demonstrated by Fig.~\ref{fig:EWC}, either an extended (model A) or
relatively small (model B) spatial extent to the emission region at
$\phi=0.994$ both produce column densities which lie between the
observed values derived by \cite{Gosset:2009} for the ``soft'' and
``hard'' hot plasma components.

Comparing the column densities from models A and B one can see that
the $E> 2\;$keV values are higher in the latter, which results from
the smaller spatial extent of the emission region at this orbital
phase in model B compared to model A (see Figs.~\ref{fig:vterm_images}
and \ref{fig:driven_images}).

\begin{figure}
  \begin{center}
    \begin{tabular}{c}
      \resizebox{80mm}{!}{\includegraphics{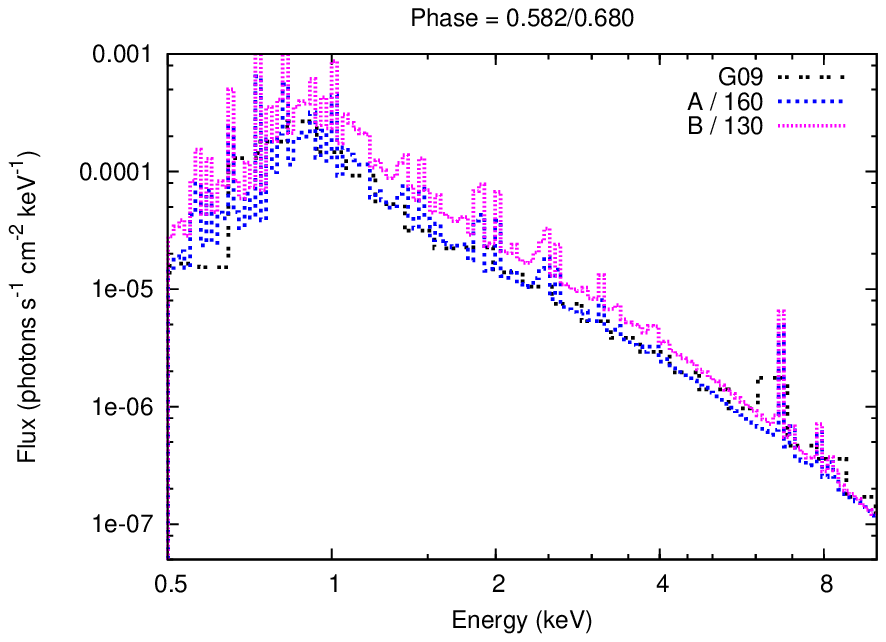}} \\
      \resizebox{80mm}{!}{\includegraphics{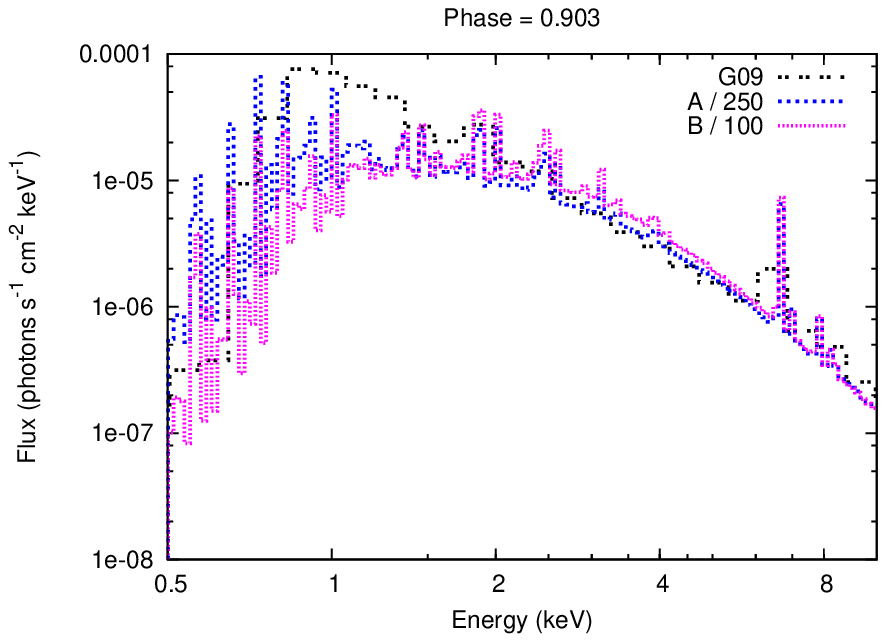}} \\
      \resizebox{80mm}{!}{\includegraphics{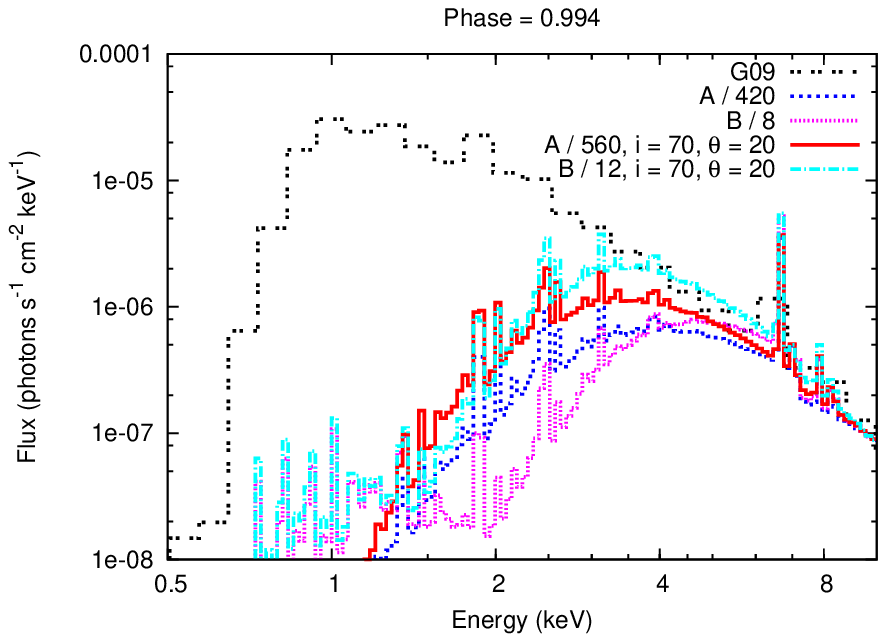}} \\ 
   \end{tabular}
    \caption{0.5-10~keV X-ray spectra calculated from models A and B
      at $\phi = 0.630\;$(upper panel), 0.903 (middle panel) and 0.994
      (lower panel). Model spectra calculated at $\phi=0.994$ using
      viewing angles of $i=70^{\circ}$ and $\theta = 20^{\circ}$ are
      also shown in the lower panel. The best-fit model spectra from
      \cite{Gosset:2009} are plotted for comparison \citep[see table 7
        of][]{Gosset:2009}. Note that the average $\phi=0.582/0.680$
      spectrum of \cite{Gosset:2009} is plotted in the upper
      panel. For illustrative purposes, the model A and B spectra have
      been rescaled by the factors noted in the plots to have an
      equivalent 10 keV flux. Note the difference in scale between the
      plots.}
    \label{fig:spectra}
  \end{center}
\end{figure}

\subsubsection{X-ray spectra}

The observed X-ray spectrum from the WCR is the product of viewing a
spatially extended region which emits at a range of energies through
an intervening medium (i.e. the line of sight absorption), which may
itself have a complicated spatial dependence. Therefore, despite the
models over-predicting the integrated 0.5-10 keV flux, a comparison of
the shape of the model spectra against observations can provide useful
insight. The X-ray spectra calculated from models A and B at
$\phi=0.63$, 0.903, and 0.994 are plotted in
Fig.~\ref{fig:spectra}. At $\phi=0.63$, a stable WCR is established in
models A and B, and consequently the X-ray spectra calculated at this
phase appear largely similar (Fig.~\ref{fig:spectra}). Both models
provide a good fit to the $\phi=0.582/0.680\;$spectrum from
\cite{Gosset:2009}, albeit with a significant overestimate in the flux
level (note that the model spectra have been rescaled to have an
equivalent 10~keV flux). As one progresses closer to periastron, the
agreement between the shape of the model spectra and observations
worsens; at $\phi=0.903$ the models underpredict the $E \simeq 1\;$keV
flux, whereas at $\phi=0.994$ the observed spectrum at $E \ltsimm
8\;$keV is poorly matched by either model\footnote{The temperature of
  the hotter plasma component is undefined in the spectral fits of
  \cite{Gosset:2009} at $\phi=0.994$. Therefore, to facilitate a
  comparison between the models and observations, we take
  $kT_{2}=4\;$keV and $Norm_{2}=1.1 \times 10^{-4}\;{\rm cm^{-5}}$
  which leads to an observed 0.5-10 keV flux of
  $\simeq1.2\times10^{-13}\;{\rm erg~s^{-1}~cm^{-2}}$, in agreement
  with, but slightly lower than, table~12 of \cite{Gosset:2009}.}. In
particular, there appears to be too much absorption at $E\ltsimm
4\;$keV close to periastron, consistent with the findings of
\cite{Gosset:2009}.

The lack of agreement between the shape of the model A and B spectra
and the \cite{Gosset:2009} spectrum at $\phi=0.994$ may be indicating
that our adopted viewing angles are incorrect and that some
adjustments are warranted. To explore this possibility tests have been
performed with viewing angles in the range $i= 70^{\circ}\;{\rm to}\;
90^{\circ}$ and $\theta = -10^{\circ}\;{\rm to}\; 40^{\circ}$. In
essence, the degree of absorption can be reduced by adopting viewing
angles which peer through less dense WR wind material. However,
although the amount of $E \ltsimm 4\;$keV emission is slightly
increased, a significant discrepancy still remains. To illustrate this
point the model A and B spectra calculated using $i=70^{\circ}$ and
$\theta=20^{\circ}$ at $\phi=0.994$ are also plotted in
Fig.~\ref{fig:spectra}; the improvement in the match is small and the
observed $E = 2\;$keV flux (in the normalized spectrum) is
underestimated by over two orders of magnitude.

The deficit in $E\ltsimm 4\;$keV flux (in the normalized spectrum) at
$\phi=0.994$ could be partially remedied by considering the ionization
state of the inner WR wind. For instance, if the gas is highly ionized
it will be less opaque. However, opposite constraints could come from
the analysis of the eclipses at optical wavelengths. In any case, an
in-depth examination of this effect is beyond the scope of the current
work.

The X-ray spectra also provide a direct probe of postshock gas
temperature, and therefore the preshock wind velocity. In particular,
the slope of the spectrum at $E \gtsimm 5\;$keV is a useful indicator
of the validity of an adopted wind velocity as X-rays at this energy
are less susceptible to absorption. Examining the spectra at
$\phi=0.582/0680$, 0.903, and 0.994 one sees that model A provides a
slightly better match to the slope of the spectrum at $E\gtsimm
5\;$keV than model B. Considering that at $\phi=0.994$ in model A the
WR wind has a preshock velocity of $1785\;{\rm km~s^{-1}}$, whereas in
model B radiative inhibition causes a slight reduction down to
$1660\;{\rm km~s^{-1}}$, this could be indicating one of two
things. Firstly, the terminal velocity of the WR wind is in fact
higher, and that at $\phi=0.994$ radiative inhibition reduces this
velocity to a preshock value of $1785\;{\rm km~s^{-1}}$. However, this
seems unlikely as the terminal wind speed of the WR wind is known with
some confidence \citep{Crowther:1995a, Hamann:2006}. Alternatively,
radiative inhibition for WR winds is not as effective as predicted by
formulations based on \cite{Castor:1975} theory
\citep[i.e.][]{Stevens:1994}. Detailed non-LTE models of WR111 support
this idea as they find the line force to be insensitive to the
gradient in the wind velocity \citep{Grafener:2005}. Furthermore, the
ionizing radiation from the O star may dramatically increase the line
force in the incoming WR wind, hence considerably enhancing the
strength of radiative braking and affecting the opening angle of the
WCR and the resulting X-ray emission \citep{Grafener:2007}.

\subsection{A revised model for WR~22} 
\label{subsec:revised_model}

The comparison between the model results and observations in the
previous sections highlights a number of discrepancies, the most
severe being a substantial overestimate of the X-ray flux across the
majority of the orbit. The major concern is the overestimate of the
2-10 keV X-ray flux as, compared to X-rays at $E< 2\;$keV, this
emission is less affected by absorption, and therefore a large
reduction in the intrinsic emission is required. In this section we
consider revisions to model parameters which may improve the agreement
with observations.

The collapse of the WCR in model B considerably improves the agreement
with the observed X-ray flux. Essentially, this occurs because the
fraction of the WR's wind being shocked is vastly reduced\footnote{By
  initiating the winds within a radius of $\sim 1.15~R_{\ast}$ we are
  artificially increasing the fraction of the WR wind being shocked by
  essentially augmenting the size of the O star. However, the
  additional X-ray flux introduced has a negligible effect on the
  results.}. As a result, a sustained collapse/disruption of the WCR
could provide a better agreement between the model and
observations. Such a situation would occur if the ram pressure of the
WR's wind massively overwhelms that of the O star's, which could be
engineered in a number of ways. Firstly, one could increase the
mass-loss rate of the WR's wind. Noting that the WR mass-loss rate
adopted in this work is relatively low in comparison to the $\simeq
4\times10^{-5}\;{\rm M_{\odot}~yr^{-1}}$ and $6.3\times10^{-5}\;{\rm
  M_{\odot}~yr^{-1}}$ determined by \cite{Crowther:1995a} and
\cite{Hamann:2006}, respectively, this would be a reasonable
alteration, but at the border of the possibilities authorized by the
eclipses in the visible domain. Alternatively, or simultaneously, one
could reduce the mass-loss rate of the O star. Considering the
uncertainty in the parameters of the O star, a factor of $\sim 2-3$
lower mass-loss rate is within reason. Repeating the calculations in
\S~\ref{subsec:estimates}, in which the winds are approximated by
$\beta$-velocity laws, we find that if the ratio of the WR and O star
mass-loss rates is increased from $\simeq 57$
(Table~\ref{tab:stellar_parameters}) to $\simeq 500$, there would be
no stable wind-wind momentum balance at any orbital phase. Considering
that a larger WR mass-loss rate may increase the amount of absorption
at periastron, and therefore potentially worsen the fit of the model
spectrum to observation, the most suitable option would appear to be a
reduction of both the WR and O star mass-loss rates, with the ratio of
the two being $\simeq 500$.

If the WR's wind ram pressure significantly overwhelms the O star's
then radiative braking could play a significant r\^{o}le
\citep{Gayley:1997}. Fig.~\ref{fig:radiative_braking} shows that,
using the stellar parameters in Table~\ref{tab:stellar_parameters},
the O star should fully brake the incoming WR wind at apastron, but
may only partially brake it at periastron. This would remain the case
if the WR's mass-loss rate is increased as proposed above. Consequently,
the relatively small increase in the observed 2-10 keV flux from
apastron to periastron may be due to the WCR transitioning from a ram
balance (supported by radiative braking) to a collision against the O
star. Curiously, the good agreement between the model spectra and
observations at $\phi=0.582/0.680$ could be indicating that the WCR
apex may not be collapsed onto the O star at this phase, perhaps
supported by effective radiative braking at phases close to
apastron. However, if a WCR were established there would either need
to be a significantly lower fraction of the WR's wind being shocked,
or much lower density postshock gas, to avoid the resulting X-ray flux
exceeding observations.

There is also the question of whether the agreement between the models
and observations could be improved if wind-clumping is
considered. Examining Fig.~\ref{fig:lc}, at $\phi\simeq0.58$ the
models overestimate the observed 2-10 keV flux by a factor of $\simeq
170$ which, as $L_{\rm X} \propto \dot{M}^2$, implies a reduction in
mass-loss rates by a factor of $\sim 13$. This greatly exceeds the
current estimates of reduction factors of 2-5 for global wind
mass-loss rates to account for clumping \citep[][]{Bouret:2003,
  Repolust:2004, Markova:2004, Fullerton:2006, Crowther:2007,
  Moffat:2008, Puls:2008}. Therefore, clumping alone cannot account
for the discrepancy between the models and observations.

In summary, the model X-ray flux will likely agree better with
observations if the WR's wind ram pressure overwhelms the O
star's. This could lead to a collapse at all phases which would
drastically reduce the observed X-ray flux (e.g. model B at phases
close to periastron in Fig.~\ref{fig:lc}). There is, however, the
intriguing possibility of a WCR established at apastron by radiative
braking transitioning to an instability-driven disruption, or collapse
of the WCR onto the O star, at periastron. For instance, the goodness
of fit of the spectral shape at apastron argues for a stable WCR
which, in the absence of a normal ram pressure balance, could be
established by radiative braking. Yet, although a collapse provides
better agreement in terms of the integrated flux, the shape of the
model B spectrum at $\phi=0.994$ does not agree very well with
observations as there is too much absorption at $E\ltsimm 2\;$keV. As
already mentioned, this may be indicating that the WR's mass-loss rate
requires a downward revision as also suggested by
\cite{Gosset:2009}. However, a minor improvement may also be attained
by considering the high ionization state, and thus lower opacity, of
the obscuring inner WR wind. Furthermore, if radiative braking is
effective at periastron then the increased opening angle of the WCR
may result in a sufficiently extended emission region such that some
$E\ltsimm 2\;$keV X-rays pass through less dense WR wind
material. Minor reductions in the model flux could also be achieved by
adopting a smaller O star radius and/or reducing the global mass-loss
rates in-line with the range of uncertainty permitted by current
estimates of wind-clumping. The former could reduce the model flux
during a WCR collapse by reducing the fraction of the WR's wind being
shocked, whereas the latter would lower the emission measure of the
postshock gas at all orbital phases.

\section{Discussion}
\label{sec:discussion}

The simulations presented in this work, although detailed and
insightful, represent an approximation to reality. In this section we
consider the potential importance of physics and/or processes which
may not be captured by our simulations, and an inherent, although
minor in this case, drawback of a numerical approach. These estimates
are by no means a substitute for further detailed investigations, and
are merely meant to serve as an indicator that there is no obvious
candidate responsible for the discrepancy between the observed and
model spectra at $\phi=0.994$ (Fig.~\ref{fig:spectra}).

\subsection{Accretion}
\label{subsec:accretion}

During periastron passage in model B, a stable WCR is not established
due to the overwhelming dominance of the WR's wind over the O star's
wind. In \S~\ref{subsec:collapse} we describe this as a collapse of
the WCR onto the O star, and it may therefore seem reasonable to
consider the possibility that the O star may accrete some of the WR's
wind. 

Capture of accreted mass will occur within the Hoyle-Lyttleton radius
\citep{Frank:2002},
\begin{equation}
  r_{\rm acc} = \frac{2 G M_{\rm O}}{v^2_{\rm \ast O} + v^2_{\rm WR}}, \label{eqn:racc}
\end{equation}
\noindent where $v_{\rm \ast O}$ and $v_{\rm WR}$ are the orbital
velocity of the O star and wind velocity of the WR wind,
respectively. Inserting values appropriate for WR22 at $\phi \simeq
1.0$, and assuming the WR wind to be at terminal velocity, we find
$r_{\rm acc}\simeq 3~{\rm R_{\odot}}$, noticeably smaller than the
adopted O star radius (Table~\ref{tab:stellar_parameters}). As $r_{\rm
  acc}$ gives an approximate value for the radius at which an incident
flow becomes gravitationally bound, this result suggests that, despite
the close proximity of the WR's wind to the O star, significant
accretion is unlikely. One may argue that if the incoming WR wind is
sufficiently radiatively braked it may lead to some capture. However,
setting $r_{\rm acc}= R_{\rm O}$ and rearranging Eq.~\ref{eqn:racc} to
determine the required O star velocity for accretion to occur it
becomes apparent that $v_{\rm WR}$ must vanish. As
Fig.~\ref{fig:radiative_braking} shows, for our adopted radiation-wind
coupling this does not appear to be the case\footnote{However, if the
  strength of the decelerative force of the O star's radiation is far
  greater \citep[c.f. ][]{Grafener:2007} then the incoming WR wind may
  be braked more effectively.}. With these points considered, we do
not anticipate that accretion will occur in WR22 during a collapse of
the WCR onto the O star at periastron passage.

\subsection{Thermal electron heat conduction}

In reality, some thermal conduction of heat across the contact
discontinuity may occur, yet its magnitude is dependent on the
magnetic field strength and orientation \citep[see
  e.g.][]{Myasnikov:1998, Orlando:2008}. If thermal electron heat
conduction were to be effective, the postshock gas temperature could
be reduced, softening the intrinsic X-ray spectrum \citep[see
  e.g.][]{Zhekov:2003}. An indication of the importance of thermal
conduction in the WCR shocks can be gained from an examination of the
thermal conduction timescale \citep{Orlando:2005},
\begin{equation}
  t_{\rm cond} = \frac{7}{2(\gamma - 1)}\frac{P_{\rm ps}}{\kappa(T_{\rm ps})(T_{\rm ps}/l^2)}~{\rm s}, \label{eqn:tcond}
\end{equation}
\noindent where $P_{\rm ps}$ and $T_{\rm ps}$ are the postshock gas
pressure and temperature, respectively, and $l$ is a characteristic
length scale (taken here to be the width of the shock at the apex of
the WCR in the simulations). Note that we assume a single temperature
for the postshock electrons and ions which should provide an accurate
approximation (c.f. \S~\ref{subsec:NEI}). The conductivity
\citep{Borkowski:1990},
\begin{equation}
  \kappa(T_{\rm ps}) = 5.6\times10^{-7} \frac{T_{\rm ps}^{5/2}}{1 + (\zeta \tan\theta)^{2}}~ {\rm erg~cm^{-1}~s^{-1}~K^{-1}},
\end{equation}
where $\zeta$ is the ratio of the number density between the cool and
hot gas and $\theta$ is the angle between the magnetic field direction
and the shock normal. Evaluating Eq.~\ref{eqn:tcond} for the case of
$\theta = 0^{\circ}$ (i.e. magnetic field lines aligned parallel to
the shock normal), and taking the value of $l$ from model B, we find
$t_{\rm cond} \simeq 4.7~{\rm days}$, which is notably longer than the
timescale for postshock gas to advect away from the apex of the WCR at
phases close periastron, $t_{\rm gas} = d/v_{\rm WR} \simeq 0.7~{\rm
  days}$, where $d$ is the binary separation. Examining the alternate
extreme, where the magnetic field lines are approximately
perpendicular to the shock normal, we find $t_{\rm cond} \simeq 40$
and 9900 days for thermal conduction between the postshock WR wind and
the preshock WR wind and the postshock O star's wind,
respectively. As a result, based on the above estimates, we do not
anticipate that thermal conduction will be largely important for
explaining the discrepancy between the observed and model spectra at
$\phi = 0.994$.

\subsection{Particle acceleration}

A fraction of the available particles could be accelerated at the
shocks to relativistic energies \citep[for a recent review see
][]{DeBecker:2007}, and in so doing the fraction of thermally emitting
particles would be reduced. The efficiency of particle acceleration in
the wind-wind collision shocks of massive star binary systems is
unclear, with estimates ranging from $\ltsimm$ a percent
\citep{Eichler:1993, Dougherty:2003, Pittard:2006} up to $\sim 50$
percent \citep{Pittardetal:2006}. If particle acceleration were to be
efficient, shock modification may occur whereby the diffusion of
non-thermal ions upstream of the subshock exert a back-pressure on the
preshock flow, causing the gas velocity to decrease prior to the
subshock \citep[see][and references
  there-in]{Pittard:2006}. Consequently, the postshock temperature
would be reduced and the observed spectrum softened. Such shock
modification is, however, unlikely in WR22 at periastron due to the
close proximity of the shocks to the O star, whereby inverse Compton
losses will limit the growth of Lorentz factors for postshock
ions. Furthermore, WR22 has been classified as a thermal radio source
\citep{Dougherty:2000}. Therefore, attributing the difference between
model and observation at $\phi=0.994$ to non-thermal effects lacks any
clear observational driver.

\subsection{Non-equilibrium ionization}
\label{subsec:NEI}

In the models presented in this work we assume that the postshock ions
and electrons are in temperature equilibrium. Interestingly, studies
of the massive star binary systems WR140 and WR147 find that the
observed soft X-ray continuum emission can naturally be explained by
non-equilibrium ionization \citep{Zhekov:2000, Pollock:2005,
  Zhekov:2007}, namely the fact that temperature equilibration between
postshock ions and electrons via Coulomb collisions may occur at some
distance downstream of the WCR apex. There are, however, two important
points which suggest that non-equilibrium ionization is not the
missing ingredient in our models of WR22. The first is that in WR140
and WR147 the shocks are thought to be collisionless. In contrast, at
periastron in WR22, the lengthscale for Coulomb collisional
dissipation, $l_{\rm i-i} \simeq 7\times10^{18} v^4_{8}/n_{i}$
\citep{Pollock:2005}, where $n_{\rm i}$ is the ion number density, is
$\simeq 4\times 10^{9}~\rm{cm}$, i.e. smaller than the finest
simulation grid cell. The second point relates to the timescale for
temperature equilibration between electrons and ions
\citep{Spitzer:1962},
\begin{equation}
  t_{\rm eq} \sim 252 \frac{\mu}{Z^2} ({\rm
    ln}~\Lambda)^{-1}\frac{T_{\rm ps}^{3/2}}{n_{\rm i}}~{\rm s}, \label{eqn:teq}
\end{equation}
\noindent where $\mu$ is the mean molecular weight, $Z$ is the
metallicity, and ln~$\Lambda$ is the Coulomb logarithm. Evaluating
Eq.\ref{eqn:teq} for postshock gas conditions in WR22 at periastron we
find $t_{\rm eq} \simeq 0.2~{\rm days}$, which implies that the
electrons and ions will rapidly equilibrate their temperatures close
to the WCR apex.

\subsection{Numerical heat conduction}

Highly radiative shocks in colliding winds binary systems are
difficult to model \citep[][]{Myasnikov:1998b, Antokhin:2004,
  Parkin_Pittard:2010}. This is mainly because of a significant
difference in length scales; the cooling length for radiative
postshock gas may become many orders of magnitude smaller than the
binary separation. Hence, sampling the cooling length of postshock gas
whilst also capturing the binary orbit is a computationally demanding
task and may affect the accuracy of the derived X-ray
spectrum. Considering the cooling length of postshock WR wind at
periastron is $\sim 6 \times 10^{12}\;$cm, and that the finest cell
size is $\simeq 6 \times 10^{10}\;$cm, there should be reasonably good
sampling. However, the large temperature and density gradient present
in the region of postshock gas in models A and B will result in some
numerical heat conduction which, as discussed in \S~\ref{sec:results},
may affect the derived X-ray spectrum
\citep{Parkin_Pittard:2010}. This situation could be improved in
future models by adopting a higher simulation resolution in regions of
large temperature and density gradients.

\section{Conclusions}
\label{sec:conclusions}

The massive WR+O-star binary system WR~22 has been investigated using
3D AMR hydrodynamic simulations which include radiative driving,
gravity, optically-thin radiative cooling, and orbital motion. Two
simulations were performed: one with instantaneously accelerated winds
(model A), and another with radiatively driven winds (model B). We
find that when the stellar winds are assumed to be instantaneously
accelerated a stable WCR is established throughout the orbit. In this
case the model massively over-predicts the observed X-ray flux,
consistent with the findings of \cite{Gosset:2009}. However, in the
second simulation which considers the acceleration of the winds, the
postshock O star's wind transits from quasi-adiabatic at apastron to
highly radiative at periastron and the character of the WCR changes
dramatically. Essentially, as the stars approach periastron the ram
pressure of the WR wind increasingly overwhelms the O star's wind,
pushing the WCR deeper into the O star's wind acceleration region, and
triggering radiative cooling in its postshock wind. The subsequent
growth of powerful NTSIs which massively disrupt the WCR is followed
by a collapse of the WCR onto the O star between $\phi\simeq
0.95-1.05$, which reduces the over-estimate of the observed flux by
the model to a factor of $\sim 6$. However, discrepancies still
remain: the observed flux is still overestimated at all orbital
phases, and the match between the models and an {\it XMM-Newton}
spectrum at $\phi=0.994$ is poor. The latter may be indicating the
necessity to consider in future models reductions in the adopted
mass-loss rates, reductions in the unshocked wind opacity due to
ionization effects, and the possibility that radiative braking is
effective at periastron.

Revisions to the adopted model parameters are considered which might
improve the agreement with observations. We conclude that the ratio of
the mass-loss rates should be increased in favour of the WR star to
the extent that a normal wind ram pressure balance does not occur at
any orbital phase, potentially leading to a sustained collapse of the
WCR onto the O star. This can be achieved with a ratio of WR to O star
mass-loss rates of $\simeq 500$, i.e. a factor of ten higher than
adopted in this work. If radiative braking is more effective than in
this work there is the interesting prospect of a WCR established at
apastron by radiative braking transitioning to an instability-driven
disruption, or collapse of the WCR onto the O star, at periastron. As
such, radiative braking may play a significant r\^{o}le for the WCR
dynamics and resulting X-ray emission.

\begin{acknowledgements}
We thank Julian Pittard and Gregor Rauw for helpful comments and a
careful reading of an earlier version of this manuscript, and the
anonymous referee for an insightful report which helped to improve
this paper. This work was supported by a PRODEX XMM/Integral contract
(Belspo). This research has made use of software which was in part
developed by the DOE-supported ASC/Alliance Center for Astrophysical
Thermonuclear Flashes at the University of Chicago.
\end{acknowledgements}


\end{document}